\documentclass[onesided]{article}
\usepackage[caption=false]{subfig}
\usepackage{xspace}
\usepackage{xparse}
\usepackage{tabularx}
\usepackage{enumitem}

\usepackage[export]{adjustbox} 
\usepackage{url}
\usepackage{wrapfig}

\newcommand{\ie}{i.e.\xspace}
\newcommand{\eg}{e.g.\xspace}
\newcommand{\etal}{et al.\xspace}

\usepackage{xparse}
\newcounter{infalg}

\NewDocumentEnvironment{informalg}{mm}{
	\refstepcounter{infalg}
	
  \setlength\extrarowheight{2pt}
	\setlist{nosep}
	\begin{table}
	\footnotesize
  \begin{tabular}{p{0.96\textwidth}}\hline
    \textbf{Algorithm \Roman{infalg}}: \textsc{#1}\\\hline
    \slshape
}{
	\\\hline
  \end{tabular}
  \end{table}
	\label{#2}
}

\newtheorem{definition}{Definition}

\newcommand{\Region}{{\mathcal R}\xspace}

\newcommand{\node}[2]{\mbox{$n_{#2}^{#1}$}\xspace}
\newcommand{\n}{\node{}{}}
\newcommand{\np}{\node{\prime}{}}

\newcommand{\edgea}[2]{\mbox{$e_{#2}^{#1}$}\xspace}
\newcommand{\edgeb}[2]{\mbox{$({#1}, {#2})$}\xspace}
\newcommand{\edgec}[3]{\mbox{${#1}:({#2}, {#3})$}\xspace}
\newcommand{\e}{\edgea{}{}}

\newcommand{\iput}[2]{\mbox{$i_{#2}^{#1}$}\xspace}
\newcommand{\ipt}{\iput{}{}}

\newcommand{\oput}[2]{\mbox{$o_{#2}^{#1}$}\xspace}
\newcommand{\opt}{\oput{}{}}

\newcommand{\argument}[2]{\mbox{$a_{#2}^{#1}$}\xspace}
\newcommand{\argmnt}{\argument{}{}}

\newcommand{\result}[2]{\mbox{$r_{#2}^{#1}$}\xspace}
\newcommand{\res}{\result{}{}}

\newcommand{\user}[2]{\mbox{$u_{#2}^{#1}$}\xspace}
\newcommand{\usr}{\user{}{}}

\newcommand{\origin}[2]{\mbox{$g_{#2}^{#1}$}\xspace}
\newcommand{\org}{\origin{}{}}

\newcommand{\card}[1]{\left\vert{#1}\right\vert}

\newcommand{\bpoint}[1]{\textbf{\textsc{#1}}}

\newcommand{\gammam}{\mbox{$\gamma$-}\xspace}
\newcommand{\gammanode}{\mbox{$\gamma$-node}\xspace}
\newcommand{\gammanodes}{\mbox{$\gamma$-nodes}\xspace}
\newcommand{\gammaregion}{\mbox{$\gamma$-region}\xspace}
\newcommand{\gammaregions}{\mbox{$\gamma$-regions}\xspace}

\newcommand{\thetam}{\mbox{$\theta$-}\xspace}
\newcommand{\thetanode}{\mbox{$\theta$-node}\xspace}
\newcommand{\thetanodes}{\mbox{$\theta$-nodes}\xspace}
\newcommand{\thetaregion}{\mbox{$\theta$-region}\xspace}
\newcommand{\thetaregions}{\mbox{$\theta$-regions}\xspace}

\newcommand{\lambdam}{\mbox{$\lambda$-}\xspace}
\newcommand{\lambdanode}{\mbox{$\lambda$-node}\xspace}
\newcommand{\lambdanodes}{\mbox{$\lambda$-nodes}\xspace}
\newcommand{\lambdaregion}{\mbox{$\lambda$-region}\xspace}
\newcommand{\lambdaregions}{\mbox{$\lambda$-regions}\xspace}

\newcommand{\deltam}{\mbox{$\delta$-}\xspace}
\newcommand{\deltanode}{\mbox{$\delta$-node}\xspace}
\newcommand{\deltanodes}{\mbox{$\delta$-nodes}\xspace}
\newcommand{\deltaregion}{\mbox{$\delta$-region}\xspace}

\newcommand{\applynode}{\mbox{$apply$-node}\xspace}
\newcommand{\applynodes}{\mbox{$apply$-nodes}\xspace}

\newcommand{\phinode}{\mbox{$\phi$-node}\xspace}
\newcommand{\phinodes}{\mbox{$\phi$-nodes}\xspace}
\newcommand{\phiregion}{\mbox{$\phi$-region}\xspace}

\newcommand{\omeganode}{\mbox{$\omega$-node}\xspace}
\newcommand{\omeganodes}{\mbox{$\omega$-nodes}\xspace}
\newcommand{\omegaregion}{\mbox{$\omega$-region}\xspace}

\newcommand{\Ozero}{\texttt{O0}\xspace}
\newcommand{\Oone}{\texttt{O1}\xspace}
\newcommand{\Otwo}{\texttt{O2}\xspace}
\newcommand{\Othree}{\texttt{O3}\xspace}
\newcommand{\Othreenovec}{\texttt{O3-no-vec}\xspace}
\newcommand{\Othreenovecstripped}{\texttt{O3-no-vec-stripped}\xspace}
\newcommand{\Os}{\texttt{Os}\xspace}

\title{RVSDG: An Intermediate Representation for Optimizing Compilers}

\usepackage{authblk}
\usepackage[scale=0.81]{geometry}

\author[1]{Nico Reissmann}
\author[1]{Jan Christian Meyer}
\author[2]{Helge Bahmann}
\author[1]{Magnus Sj\"alander}

\affil[1]{Norwegian University of Science and Technology}
\affil[2]{Auterion AG}

\date{}

\begin{document}
\maketitle
\begin{abstract}
  Intermediate Representations (IRs) are central to optimizing compilers as the
  way the program is represented may enhance or limit analyses and %
  transformations. %
  Suitable IRs focus on exposing the most relevant information and establish
  invariants that different compiler passes can rely on. %
  While control-flow centric IRs appear to be a natural fit for imperative
  programming languages, analyses required by compilers have increasingly
  shifted to understand data dependencies and work at multiple abstraction
	layers at the same time. %
  This is partially evidenced in recent developments such as the MLIR proposed
  by Google. %
  However, rigorous use of data flow centric IRs in general purpose compilers
  has not been evaluated for feasibility and usability as previous works provide
  no practical implementations. %

  We present the Regionalized Value State Dependence Graph (RVSDG) IR for
  optimizing compilers. %
  The RVSDG is a data flow centric IR where nodes represent computations, edges
  represent computational dependencies, and regions capture the hierarchical
  structure of programs. %
  It represents programs in demand-dependence form, implicitly supports
  structured control flow, and models entire programs within a single IR. %
  We provide a complete specification of the RVSDG, construction and destruction
  methods, as well as exemplify its utility by presenting Dead Node and Common
  Node Elimination optimizations. %
  We implemented a prototype compiler and evaluate it in terms of performance,
  code size, compilation time, and representational overhead. %
  Our results indicate that the RVSDG can serve as a competitive IR in
  optimizing compilers while reducing complexity.
\end{abstract}

\section{Introduction}

Intermediate representations (IRs) are at the heart of every modern compiler. These data structures
represent programs throughout compilation, connect individual compiler stages, and provide
abstractions to facilitate the implementation of analyses, optimizations, and program
transformations. A suitable IR highlights and exposes program properties that are important to the
transformations in a specific compiler stage. This reduces the complexity of optimizations and
simplifies their implementation.

Modern embedded systems have become increasingly parallel %
as system designers strive to improve their computational power and energy
efficiency. %
Increasing the number of cores in a system enables each core to be operated at a
lower clock frequency and supply voltage, improving
overall energy efficiency while providing sufficient system performance. %
Multi-core systems also reduce the total system cost by enabling the consolidation
of multiple functionalities onto a single chip.
In order to take full advantage of these systems, optimizing compilers need to
expose a program's available parallelism. %
This has led to an interest in developing more efficient program
representations~\cite{Cordes10,Lattner20}
and methodologies and frameworks~\cite{Castrillon13} for exposing the necessary information.

Data flow centric IRs, such as the Value (State) Dependence Graph
(V(S)DG)~\cite{Weise94, Johnson03, Johnson04, Lawrence07, Stanier211, Stanier12}, show promises as a
new class of IRs for optimizing compilers. %
These IRs are based on the observation that many optimizations require data flow
rather than control flow information, and shift the focus to explicitly expose
data instead of control flow. %
They represent programs in demand-dependence form, encode structured control
flow, and explicitly model data flow between operations. %
This raises the IR's abstraction level, permits simple and powerful
implementations of data flow optimizations, and helps to expose the inherent
parallelism in programs~\cite{Lawrence07, Johnson04, Stanier211}. %
However, the shift in focus from explicit control flow to only structured and
implicit control flow requires more sophisticated construction and destruction
methods~\cite{Weise94, Lawrence07, Stanier12}. %
In this context, Bahmann~\etal~\cite{Bahmann15} presents the \emph{Regionalized Value State
Dependence Graph} (RVSDG) and conclusively addresses the problem of intra-procedural control
flow recovery for demand-dependence graphs. %
They show that the RVSDG's restricted control flow constructs do not limit the complexity of
the recoverable control flow. %

In this work, we are concerned with the aspects of unified program
representation in the RVSDG. We present the required RVSDG constructs, consider
construction and destruction at the program level, and show feasibility and
practicality of this IR for optimizations by providing a practical compiler implementation.
Specifically, we make the following contributions: \emph{i)} A complete RVSDG specification,
including intra- and inter-procedural constructs. \emph{ii)} A complete description of RVSDG
construction and destruction, augmenting the previously proposed algorithms with the construction
and destruction of inter-procedural constructs, as well as the handling of intra-procedural
dependencies during construction. \emph{iii)} A presentation of Dead Node Elimination (DNE) and
Common Node Elimination (CNE) optimizations to demonstrate the RVSDG's utility. DNE combines dead
and unreachable code elimination, as well as dead function removal. CNE permits the removal of
redundant computations by detecting congruent operations. \emph{iv)} A publicly
available~\cite{jlm} prototype compiler that implements the discussed concepts.  It consumes and
produces LLVM IR, and is to our knowledge the first optimizing compiler that uses a demand
dependence graph as IR. \emph{v)} An evaluation of the RVSDG in terms of performance and size of
the produced code, as well as compile time and representational overhead.

Our results show that the RVSDG can serve as the IR in a compiler's optimization stage, producing
competitive code even with a conservative modeling of programs using a single memory state.
Even though this leaves significant parallelization potential
unused, it already yields satisfactory results compared to control-flow
based approaches.
This work paves the way for %
further exploration of the RVSDG's properties and their effect on optimizations
and analyses, as well as its usability in code generation for dataflow and
parallel architectures.

\section{Motivation}\label{sec:motivation}

Contemporary optimizing compilers are mainly based on variants of the control flow
graph as imperative program representations. %
These representations preserve sequential execution semantics of the input program, such as access order of aliased references.
The LLVM representation is based on the instruction set
of a virtual CPU with operation semantics resembling real CPUs. %
This choice of representation is somewhat at odds with the requirements of code
optimization analysis, which often focuses on {\em data dependence} instead.
As Table~\ref{tab:llvm-opts} shows, most executed optimization passes
are concerned with data flow analysis in the form of SSA construction
and interpretation, or in-memory data structures in the form of alias analysis
and/or memory SSA. %

We propose the data-dependence centric RVSDG as an alternative. %
While it requires more effort to construct the RVSDG from imperative
programs and recover control flows for code generation, we believe this cost
is more than recovered by benefits to analyses and optimizations.
The following sections provide illustrative examples.

\begin{wraptable}{l}{0.5\textwidth}
  \vspace{-1em}
	\caption{Thirteen most invoked LLVM 7.0.1 passes at \texttt{O3}.\label{tab:llvm-opts}}
	\resizebox{0.5\textwidth}{!}{%
	\begin{tabular}{l|c}
		Optimization                                                             & \# Invocations\\\hline\hline
		1. Alias Analysis (\texttt{-aa})                                         & 19\\
		2. Basic Alias Analysis (\texttt{-basicaa})                              & 18\\
		3. Optimization Remark Emitter (\texttt{-opt-remark-emitter})            & 15\\
		4. \textbf{Natural Loop Information (\texttt{-loops})}                   & 14\\
		5. Lazy Branch Probability Analysis (\texttt{-lazy-branch-prob})         & 14\\
		6. Lazy Block Frequency Analysis (\texttt{-lazy-block-freq})             & 14\\
		7. \textbf{Dominator Tree Construction (\texttt{-domtree})}              & 13\\
		8. Scalar Evolution Analysis (\texttt{-scalar-evolution})                & 10\\
		9. \textbf{CFG Simplifier (\texttt{-simplifycfg})}                       & 8\\
		10. Redundant Instruction Combinator (\texttt{-instcombine})             & 8\\
		11. \textbf{Natural Loop Canonicalization (\texttt{-loop-simplify})}     & 8\\
		12. \textbf{Loop-Closed SSA Form (\texttt{-lcssa})}                      & 7\\
		13. \textbf{Loop-Closed SSA Form Verifier (\texttt{-lcssa-verification})}& 7\\\hline
		Total                                                                    & 155\\\hline
		\textbf{SSA Restoration}                                                 & 14\\
	\end{tabular}
      }
  \vspace{-1em}
\end{wraptable}

\begin{figure*}
	\subfloat[Code\label{fig:exp1-code}]{
		\includegraphics[scale=0.95]{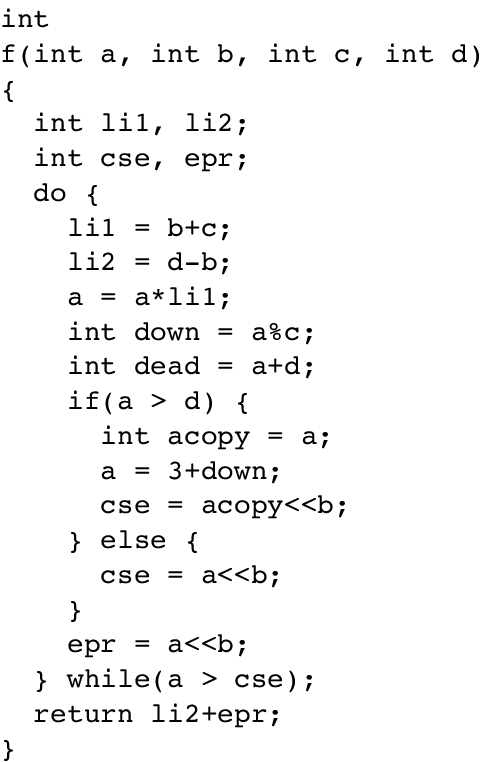}
	}
	\hfill
	\subfloat[CFG in SSA form\label{fig:exp1-cfg}]{
		\includegraphics[scale=0.95]{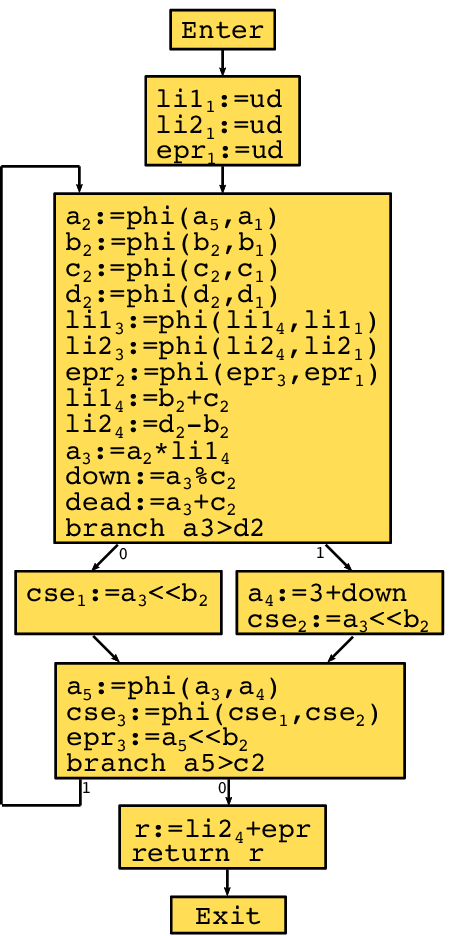}
	}
	\hfill
	\subfloat[Unoptimized RVSDG\label{fig:exp1-rvsdg}]{
		\includegraphics[scale=0.95]{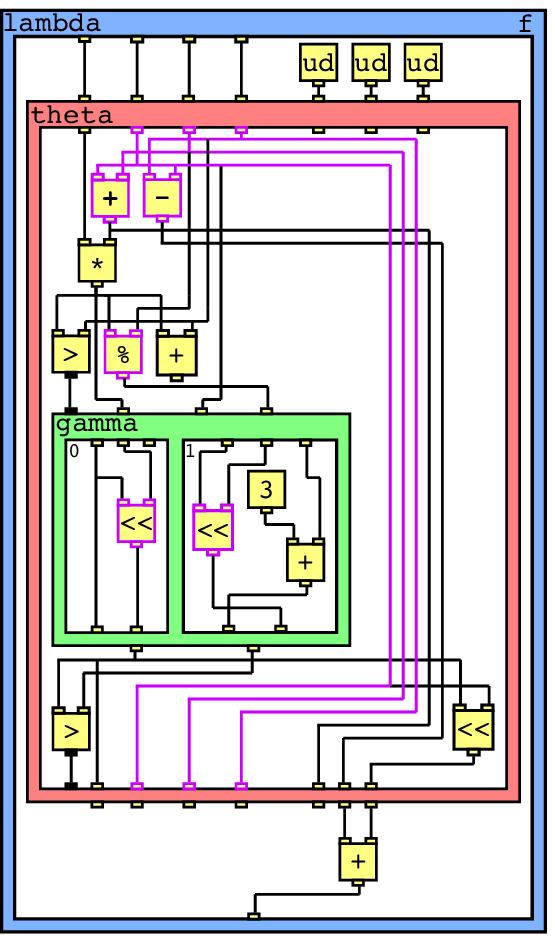}
	}

	\subfloat[Optimized RVSDG\label{fig:exp1-rvsdg-opt}]{
		\includegraphics[scale=0.95]{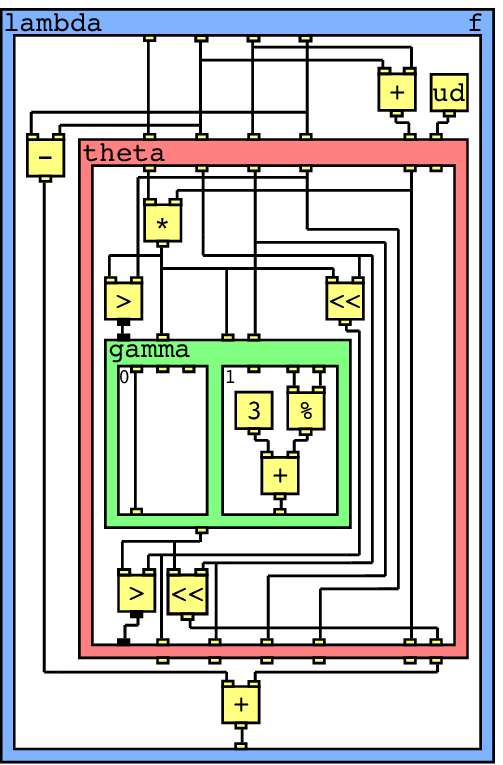}
	}
	\hfill
	\subfloat[Code\label{fig:exp2-code}]{
		\includegraphics[scale=0.95]{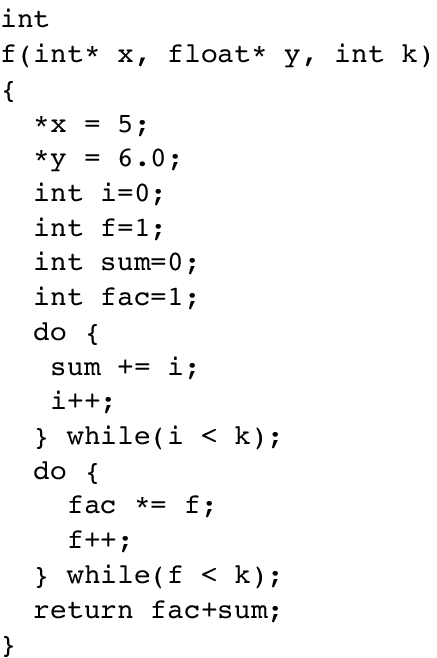}
	}
	\hfill
	\subfloat[RVSDG of Code~\ref{fig:exp2-code}\label{fig:exp2-rvsdg}]{
		\includegraphics[scale=0.95]{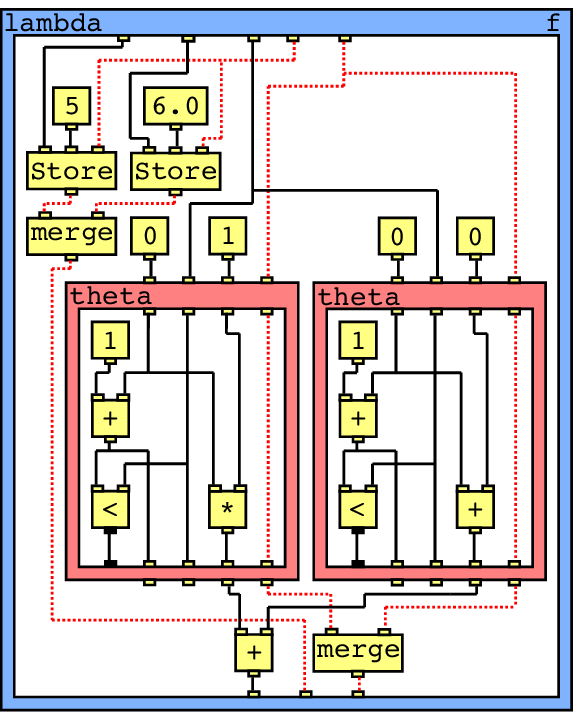}
	}
	\caption{RVSDG Examples}
	\label{fig:exp1}
\end{figure*}

\subsection{Simplified Compilation by Strong Representation Invariants}\label{sec:rep-invariants}

The Control Flow Graph (CFG) in Static Single Assignment (SSA)
form~\cite{Cytron91} is the dominant IR for optimizations in modern
imperative language compilers~\cite{Stanier13}. %
Its nodes represent a list of totally ordered operations, and its edges a
program's possible control flow paths, permitting efficient control flow
optimizations and simple code generation. %
The CFG's translation to SSA form improves the efficiency of many data flow
optimizations~\cite{Rosen88,Wegman91}. %
Figure~\ref{fig:exp1-code} shows a function with a simple loop and a
conditional, and Figure~\ref{fig:exp1-cfg} shows the
corresponding~CFG~in~SSA~form. %

SSA form is not an intrinsic property of the CFG, but a
specialized variant that must be actively maintained. %
Compiler passes such as jump threading or live-range splitting may perform
transformations that cause the CFG to no longer satisfy this form. %
As shown in Table~\ref{tab:llvm-opts}, LLVM requires SSA restoration~\cite{Choi96} in 14 different
passes.

Moreover, CFG-based compilers must frequently (re-)discover and canonicalize loops,
or establish various invariants besides SSA form. %
Table~\ref{tab:llvm-opts} shows that six of the 13 most invoked passes in LLVM
only perform such tasks, and account for 21\% of all invocations. %
This lack of invariants complicates implementation of optimizations
and analyses, increases engineering effort, prolongs compilation
time, and leads to compiler bugs~\cite{llvm-bug1,llvm-bug2,llvm-bug3}. %

In contrast, the RVSDG is always in strict SSA form as edges connect each operand
input to only one output. It explicitly exposes program structures such as loops
in a tree structure (Section~\ref{sec:rvsdg}), similarly to the Program Structure
Tree~\cite{Johnson94}. This makes SSA restoration and the other helper passes
from Table~\ref{tab:llvm-opts} redundant.
Figure \ref{fig:exp1-rvsdg} shows the RVSDG corresponding
to Figure \ref{fig:exp1-code}. %
It is an acyclic demand-dependence graph where nodes represent simple operations
or control flow constructs, and edges represent dependencies between
computations (Section~\ref{sec:rvsdg}). %
In Figure \ref{fig:exp1-rvsdg}, simple operations are colored yellow,
conditionals are green, loops are red, and functions are blue. %

\subsection{Unified Representation of Different Levels of Program Structures}

While the CFG can represent %
a single procedure, representation of programs as a whole requires additional
data structures such as call graphs. %
The RVSDG can represent a program as a unified data structure where a def-use
dependency of one function on another is modeled the same way as the def-use
dependency of scalar quantities. %
This makes it possible to apply the same transformation at multiple levels,
and considerably reduce the number of transformation passes and
algorithms, \eg, by uniform treatment of unreachable code, dead function analysis,
 and dead variable analysis (Section~\ref{sec:dne}).

\subsection{Strongly Normalized Representation}

The RVSDG program representation is much more strongly normalized than control flow
representations. %
Programs differing only in the ordering of (independent) operations result in
the same RVSDG representation, while state edges ensure the correct evaluation order
of stateful computations. Loops and conditionals always take a single
canonical form. %
These normalizations already simplify the implementation of
transformations~\cite{Weise94,Johnson04,Lawrence07} and eliminate the need for (repeated)
compiler analysis passes such as loop detection. %

Some common program optimizing transformations take a particular simple form in
the RVSDG representation. %
For example, Figure~\ref{fig:exp1-rvsdg-opt} shows the optimized RVSDG of
Figure~\ref{fig:exp1-rvsdg}, illustrating some of these optimizations: %
The inputs to the ``upper left'' plus operation are easily recognized as loop
invariant because their ``loop entry ports'' connect directly to the
corresponding ``loop exit ports'' (operations, ports, and edges highlighted in
purple). %
A simple push strategy allows to recursively identify data dependent operations
as invariant and hoist them out of the loop: %
The addition and subtraction computing \verb;li1; and \verb;li2; are moved out
of the loop (theta) as their operands, \ie, \verb;b;, \verb;c;, and \verb;d;, are
loop invariant (all three of them connect the entry of the loop to the exit). %
Similarly, the shift operation common to both conditional branches is hoisted
and combined, while the division operation is moved into the conditional as it
is only used in one alternative. %
In contrast to CFG-based compilers, all these optimizations are performed
directly on the unoptimized RVSDG of Figure~\ref{fig:exp1-rvsdg} and can be
performed in a single regular pass. %
No additional data structures or helper passes are required. %
See also Section~\ref{sec:optimizations} for further details. %

\subsection{Exposing Independent Computations}

CFGs implicitly represent a single global machine state by sequencing every 
operation that affects it. While RVSDG can model the same machine,
it is not limited to this interpretation. %
The RVSDG can also model systems consisting of multiple
{\em independent} states. Figures~\ref{fig:exp2-code},\ref{fig:exp2-rvsdg}
illustrate this concept with a function that contains two non-aliasing store
operations (targeting memory objects of incompatible types) and two
independent loops.

In a CFG, both stores and loops are strictly ordered. Their
mutual independence needs to be established by explicit compiler passes (and
may need to be re-established multiple times during the compilation
process as the number of alias analysis passes in Table~\ref{tab:llvm-opts} illustrate)
and represented using auxiliary data structures and/or annotations. In
contrast, the RVSDG permits the encoding of such information directly in the graph, as shown in
Figure~\ref{fig:exp2-rvsdg}. Disjoint memory regions (consisting of {\tt int}-typed
and {\tt float}-typed memory objects) are modeled as disjoint states, exposing
the independence of affecting operations in the representation. RVSDG can
in principle go even further in representing a {\em memory SSA} form that is
not formally any different from {\em value SSA} form, enabling the same kind
of optimizations to be applied to both.

\subsection{Multiple levels of abstraction}

The RVSDG is fairly abstract in nature and can contain operational nodes at vastly
different abstraction levels: Operational nodes may closely correspond to ``source
code'' level constructs operating on data structures modeled as state, or may
map to individual machine instructions affecting machine and memory state. This
offers an opportunity to structure compilers in a way that can preserve
considerably more source code semantics and utilize it at any later stage in the
translation. For example, contents of two distinct {\tt std::vector} instances
can never alias by language semantics, but this fact is entirely lost on
present-day compilers due to early lowering into a machine-like representation
without the capability to convey such high-level semantics. The RVSDG does not face
any such limitations (vector contents could be modeled as independent states
from the very beginning), can optimize at multiple levels of abstraction,
and can preserve vital invariants across abstraction levels. We expect this effect to become
particularly pronounced the more input programs are formulated above the abstraction level
of the C language, \eg, functional languages or languages expressing contracts
on affected state.

\subsection{Summary}

The RVSDG raises the IR abstraction level by enforcing desirable properties, such as
SSA form, explicitly encoding important structures, such as loops, and relaxing the overly strict
order of the input program. This leads to a more normalized program representation and avoids
many idiosyncrasies and artifacts from other IRs, such as the CFG, and further helps to expose
parallelism in programs.

\section{Related Work}\label{sec:relwork}

A cornucopia of IRs has been presented in the literature to better expose desirable
program properties for optimizations. For brevity, we restrict our discussion to the
most prominent IRs, only highlighting their strengths and weaknesses in comparison to
the RVSDG, and refer the reader to Stanier~\etal~\cite{Stanier13} for a greater survey.

\subsection{Control (Data) Flow Graph}
The Control Flow Graph (CFG)~\cite{Allen70} exposes the intra-procedural control flow of a
function. Its nodes represent basic blocks, \ie, an ordered list of operations without
branches or branch targets, and its edges represent the possible control flow paths between these
nodes. This explicit exposure of control flow simplifies certain analyses, such as loop
identification or irreducibility detection, and enables simple target code generation. The CFG's
translation to SSA form~\cite{Cytron91}, or one of its variants, such as gated SSA~\cite{Tu95},
thinned gated SSA~\cite{Havlak93}, or future gated SSA~\cite{Ding14}, additionally improves the
efficiency of data flow optimizations~\cite{Wegman91,Rosen88}. These properties along with its
simple construction from a language's abstract syntax tree made the CFG in SSA form the predominant
IR for imperative language compilers~\cite{Stanier13}, such as LLVM~\cite{Lattner04} and
GCC~\cite{gcc}. However, the CFG has also been criticized as an IR for optimizing
compilers~\cite{Ferrante87,Johnson03,Johnson04,Lawrence07,Weise94,Zaidi15,Zaidi15-2}:

\begin{enumerate}
	\item It is incapable of representing inter-procedural information. It requires additional IRs,
		\eg, the call graph, to represent such information.

	\item It provides no structural information about a procedure's body. Important structures, such
		as loops, and their nesting needs to be constantly (re-)discovered for optimizations, as well
		as normalized to make them amenable for transformations.

	\item It emphasizes control dependencies, even though many optimizations are based on the flow
		of data. This is somewhat mitigated by translating it to SSA form or one of its variants, but
		in turn requires SSA restoration passes~\cite{Choi96} to ensure SSA invariants.

	\item It is an inherently sequential IR. The operations in basic blocks are listed in a
		sequential order, even if they are not dependent on each other. Moreover, this
		sequentialization also exists for structures such as loops, as two independent loops can only
		be represented in sequential order. Thus, the CFG is by design incapable of explicitly
		encoding independent operations.

	\item It provides no means to encode additional dependencies other than control and true data
		dependencies. Other information, such as loop-carried dependencies or alias information, must
		regularly be recomputed and/or memoized in addition to the CFG.
\end{enumerate}

The Control Data Flow Graph (CDFG)~\cite{Namballa04} tries to mitigate the sequential nature of
the CFG by replacing the sequence of operations in basic blocks with the Data Flow
Graph (DFG)~\cite{Dennis80}, an acyclic graph that represents the flow of data between operations.
This relaxes the strict ordering within a basic block, but does not expose instruction level
parallelism beyond basic block boundaries or between program structures.

\subsection{Program Dependence Graph/Web}
The Program Dependence Graph (PDG) \cite{Ferrante87,Horwitz288} combines control and data flow
within a single representation. It features data and control flow edges, as well as statement,
predicate, and region nodes. Statement nodes represent operations, predicate nodes represent
conditional choices, and region nodes group nodes with the same control dependency. If a
region's control dependencies are fulfilled, then its children can be executed in
parallel. Horwitz \etal~\cite{Horwitz88} extended the PDG to model inter-procedural dependencies
by incorporating procedures into the graph.

The PDG improves upon the CFG by employing region nodes to relax the overly restrictive sequence
of operations. This relaxed sequence combined with the unified representation of data and control
dependencies simplifies complex optimizations, such as code vectorization~\cite{Baxter89} or the
extraction of thread-level parallelism~\cite{Ottoni05,Sarkar91}. However, the unified data and
control flow representation results in a large number of edge types, five in Ferrante
\etal~\cite{Ferrante87} and four in Horwitz \etal~\cite{Horwitz288}, which need to be maintained
to ensure the graph's invariants. The PDG suffers from aliasing and side-effect problems, as it
supports no clear distinction between data held in register and memory. This complicates or can
even preclude its construction altogether~\cite{Johnson04}. Moreover, program structure and SSA
form still need to be discovered and maintained.

The Program Dependence Web (PDW)~\cite{Ottenstein90} extends the PDG and gated SSA~\cite{Tu95}
to provide a unified representation for the interpretation of programs using control-, data-, or
demand-driven execution models. This simplifies the mapping of programs written in different
paradigms, such as the imperative or functional paradigm, to different architectures, such as
Von-Neumann and dataflow architectures. In addition to the elements of the PDG, the PDW adds $\mu$
nodes to manage initial and loop-carried values and $\eta$ nodes to manage loop-exit
values. Campbell \etal~\cite{Campbell93} further refined the definition of the PDW by replacing
$\mu$ nodes with $\beta$ nodes and eliminating $\eta$ nodes. As the PDW is based on the PDG, it
suffers from the same aliasing and side-effect problems. PDW's additional constructs
further complicate graph maintenance and its construction is elaborate, requiring three
additional passes over a PDG, and is limited to programs with reducible control flow.

\subsection{Value (State) Dependence Graph}
The Value Dependence Graph (VDG)~\cite{Weise94} abandons the explicit representation of control
flow and only models the flow of values using ports. Its nodes represent simple operations, the
selection between values, or functions, using recursive functions to model loops.
The VDG is implicitly in SSA form and abandons the sequential order of operations from the CFG, as
each node is only dependent on its values. However, modeling only data flow between stateful
computations raises a significant problem in terms of preservation of program semantics, as the
"evaluation of the VDG may terminate even if the original program would not..."~\cite{Weise94}.

The Value State Dependence Graph (VSDG)~\cite{Johnson03,Johnson04} addresses the VDG's termination
problem by introducing state edges. These edges are used to model the sequential
execution of stateful computations. In addition to nodes for representing simple operations
and selection, it introduces nodes to explicitly represent loops. Like the VDG, the VSDG is
implicitly in SSA form, and nodes are solely dependent on required operands, avoiding a sequential
order of operations. However, the VSDG supports no inter-procedural constructs, and its selection
operator is only capable of selecting between two values based on a predicate. This complicates
destruction, as selection nodes must be combined to express conditionals. Even worse, the VSDG
represents all nodes as a flat graph, which simplifies optimizations~\cite{Johnson04}, but has a
severe effect on evaluation semantics. Operations with side-effects are no longer guarded by
predicates, and care must be taken to avoid duplicated evaluation of these operations. In fact,
for graphs with stateful computations, lazy evaluation is the only safe strategy~\cite{Lawrence07}.
The restoration of a program with an eager evaluation semantics complicates destruction immensely,
and requires a detour over the PDG to arrive at a unique CFG~\cite{Lawrence07}.
Zaidi \etal~\cite{Zaidi15-2,Zaidi15} adapted the VSDG to spatial hardware and sidestepped this
problem by introducing a predication-based eager/dataflow semantics. The idea is to effectively
enforce correct evaluation of operations with side-effects by using predication. While this seems
to circumvent the problem for spatial hardware, it is unclear what the performance implications
would be for conventional~processors.

The RVSDG solves the VSDG's eager evaluation problem by introducing regions. %
These regions enable the modeling of control flow constructs as nested nodes, and the guarding of
operations with side-effects. This avoids any possibility of duplicated evaluation, and in turn
simplifies RVSDG destruction. Moreover, nested nodes permit the explicit encoding of a program's
hierarchical structure into the graph, further simplifying optimizations.

\section{The Regionalized Value State Dependence Graph}\label{sec:rvsdg}
A Regionalized Value State Dependence Graph (RVSDG) is an acyclic hierarchical multigraph
consisting of nested regions. A region $\Region = (A, N, E, R)$ represents a computation with
argument tuple $A$, nodes $N$, edges $E$, and result tuple $R$, as illustrated
in Figure~\ref{fig:notation}. A node can be either
\emph{simple}, \ie, it represents a primitive operation, or \emph{structural}, \ie, it
contains regions. Each node $\n \in N$ has a tuple of inputs $I$ and outputs $O$. In case of
simple nodes, they correspond to arguments and results of the represented operation, whereas for
structural nodes, they map to arguments and results of the contained regions. For nodes
$\node{}{1}, \node{}{2} \in N$, an edge $\edgeb{\org}{\usr} \in E$ connects either output
$\org \in O_{n_1}$ or argument $\org \in A$ to either input $\usr \in I_{n_2}$ or result
$\usr \in R$ of matching type. We refer to \org as the \emph{origin} of an edge, and to \usr as the
\emph{user} of an edge. Every input or result is the user of \emph{exactly one} edge, whereas
outputs or arguments can be the origins of multiple edges. All inputs or results of an origin are
called its \emph{users}. The corresponding node of an origin is called its \emph{producer}, whereas
the corresponding node of a user is called \emph{consumer}. Correspondingly, the set of nodes of
all users of an origin are referred to as its \emph{consumers}. The types of inputs and outputs
are either \emph{values}, representing arguments or results of computations, or \emph{states}, used
to impose an order on operations with side-effects. A node's \emph{signature} are the
types of its inputs and outputs, whereas a region's signature are the types of its arguments
and results. %
Throughout this paper, we use \n, \e, \ipt, \opt, \argmnt, and \res with
sub- and superscripts to denote individual \textbf{n}odes, \textbf{e}dges, \textbf{i}nputs,
\textbf{o}utputs, \textbf{a}rguments, and \textbf{r}esults, respectively. We use \org and \usr to
denote an edge's ori\textbf{g}in and \textbf{u}ser, respectively. An edge \e from origin \org to
user \usr is also denoted as \edgec{\e}{\org}{\usr}, or short \edgeb{\org}{\usr}.

The RVSDG can model programs at different abstraction levels. It can represent simple
data-flow graphs such as those used in machine learning frameworks, but it can also represent
programs at the machine level as used in compiler back-ends for code generation. This
flexibility makes it possible to use the RVSDG for the entire compilation pipeline. In this paper,
we target an abstraction level similar to that of LLVM IR. This permits us to illustrate all of
the RVSDG's features without involving architecture-specific details. The rest of this section
defines the necessary constructs.

\subsection{Intra-procedural Nodes}\label{sec:intra-nodes}

This section defines the nodes for representing the intra-procedural aspects of programs. It
explains simple nodes and discusses the two structural nodes required for modeling intra-procedural
program behavior:

\begin{enumerate}
	\item \emph{Gamma-Nodes} model conditionals with symmetric split and joins, such as
		\texttt{if-then-else} statements.
	\item \emph{Theta-Nodes} represent tail-controlled loops, \ie, \texttt{do-while} loops.
\end{enumerate}

\subsubsection{Simple nodes}
\begin{figure*}
	\subfloat[Notation\label{fig:notation}]{
			\includegraphics[scale=0.84]{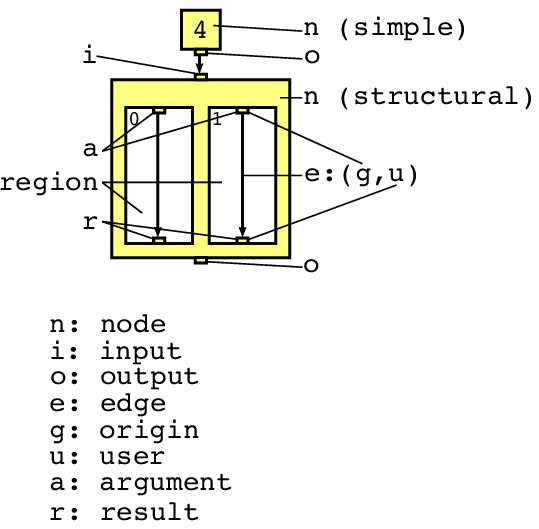}
	}
	\hfill
	\subfloat[Simple nodes\label{fig:expr}]{
			\includegraphics[scale=0.84]{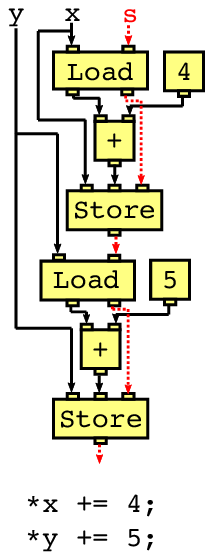}
	}
	\hfill
	\subfloat[\gammanode\label{fig:gamma}]{
			\includegraphics[scale=0.84]{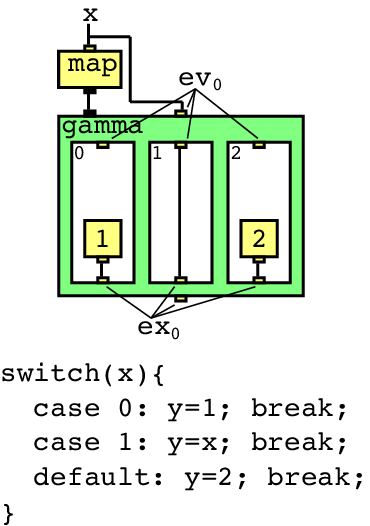}
	}
	\hfill
	\subfloat[\thetanode\label{fig:theta}]{
			\includegraphics[scale=0.84]{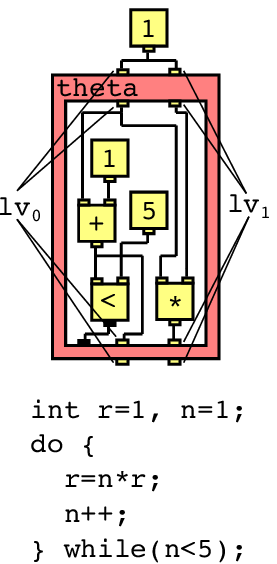}
	}
	\caption{Notation as well as examples for the usage of simple, \gammam and \thetanodes.}
	\label{fig:example}
\end{figure*}

Simple nodes model primitive operations such as addition, subtraction, load, and store. They have
an operator associated with them, and a node's signature must correspond to the signature of
its operator. Simple nodes map their input value tuple to their output value tuple by evaluating
their operator with the inputs as arguments, and associating the results with their outputs.
Figure \ref{fig:expr} illustrates the use of simple nodes as well as value and state edges. Solid
lines represent value edges, whereas dashed lines represent state edges. Nodes have as many
value inputs and outputs as their corresponding operations demand. The ordering of the load and
store nodes is preserved by sequentializing them with the help of state edges.

\subsubsection{Gamma-Nodes}
A \gammanode models a decision point and contains regions
$\Region_0, ..., \Region_k \mid k > 0$ of matching signature. Its first input is a
\emph{predicate}, which determines the region under evaluation. It evaluates to an
integer $v$ with $0 \le v \le k$. The values of all other inputs are mapped to the corresponding
arguments of region $\Region_v$, $\Region_v$ is evaluated, and the values of its results are mapped
to the outputs of the \gammanode.

\gammanodes represent conditionals with symmetric control flow splits and
joins, such as \verb;if-then-else; or \verb;switch; statements without fall-throughs. Figure
\ref{fig:gamma} shows a \gammanode. It contains three regions: one for each case, and a default
region. The map node takes the value of $x$ as input and maps it to zero, one, or two, determining
the region under evaluation. This region is evaluated and its result is mapped
to the \gammanode's output.

We define the \emph{entry variable} of a \gammanode as a pair of an input and the arguments the
input maps to during evaluation, as well as the \emph{exit variable} of a \gammanode as a pair of
an output and the results the output could receive its value from:

\begin{definition}
The pair $ev_{l} = (\iput{}{l+1}, A_{l})$ is the $l$-th \emph{entry variable} of a \gammanode with
$k+1$ regions. It consists of the $l+1$-th input and tuple $A_{l} = \{a^{\Region_0}_{l}, ...,
a^{\Region_k}_{l}\}$ with the $l$-th argument from each region. We refer to the set of all
entry variables as $EV$.
\end{definition}

\begin{definition}
The pair $ex_{l} = (R_l, \oput{}{l})$ is the $l$-th \emph{exit variable} of a \gammanode with $k+1$
regions. It consists of a tuple $R_l = \{r^{\Region_0}_l, ...,r^{\Region_k}_l\}$ of the $l$-th
result from each region and the $l$-th output they would map to. We refer to the set of all exit
variables as $EX$.
\end{definition}

Figure~\ref{fig:gamma} shows the \gammanode's only entry and exit variable annotated.

\subsubsection{Theta-Nodes}
A \thetanode models a tail-controlled loop. It contains one
region that represents the loop body. The length and signature of its input tuple equals that of
its output, or the region's argument tuple. The first region result is a \emph{predicate}.
Its value determines the continuation of the loop. When a \thetanode is evaluated, the values of
all its inputs are mapped to the corresponding region arguments and the body is evaluated. When
the predicate is true, all other results are mapped to the corresponding arguments for the next
iteration. Otherwise, the result values are mapped to the corresponding outputs. The loop
body of an iteration is always fully evaluated before the evaluation of the next iteration.
This avoids ``deadlock`` problems between computations of the loop body and the predicate, and
results in well-defined behavior for non-terminating loops that update external state.

\thetanodes permit the representation of \verb;do-while; loops. In combination with \gammanodes,
it is possible to model head-controlled loops, \ie, \verb;for; and \verb;while; loops. Thus,
employing tail-controlled loops as basic loop construct enables us to express more complex loops as
a combination of basic constructs. This normalizes the representation and reduces the complexity
of optimizations as there exists only one construct for loops. Another benefit of tail-controlled
loops is that their body is guaranteed to execute at least once, enabling the unconditional
hoisting of invariant code with side-effects.

Figure \ref{fig:theta} shows a \thetanode with two loop variables, $n$ and $r$, and an additional
result for the predicate. When the predicate evaluates to true, the results for $n$ and
$r$ of the current iteration are mapped to the region arguments to continue with the next
iteration. When the predicate evaluates to false, the loop exits and the results are mapped to the
node's outputs. We define a \emph{loop variable} as a quadruple that represents a value routed
through a \thetanode:

\begin{definition}
The quadruple $lv_{l} = (\iput{}{l}, \argument{}{l}, \result{}{l+1}, \oput{}{l})$ is the $l$-th
\emph{loop variable} of a \thetanode. It consists of the $l$-th input \iput{}{l}, argument
\argument{}{l}, and output \oput{}{l}, and the $l+1$-th result of a \thetanode. We refer to
the set of all loop variables as $LV$.
\end{definition}

Figure~\ref{fig:theta} shows the \thetanode's two loop variables annotated.

\subsection{Inter-procedural Nodes}\label{sec:inter-nodes}
This section defines the four structural nodes used for modeling the inter-procedural aspects of
programs:

\begin{enumerate}
	\item \emph{Lambda-Nodes} are used for modeling procedures and functions.
	\item \emph{Delta-Nodes} model global variables.
	\item \emph{Phi-Nodes} represent mutually recursive environments, such as (mutually) recursive
		functions.
	\item \emph{Omega-Nodes} represent translation units.
\end{enumerate}

\subsubsection{Lambda-Nodes}
A \lambdanode models a function and contains a single region representing a
function's body. It features a tuple of inputs and a single output. The inputs refer to external
variables the \lambdanode depends on, and the output represents the \lambdanode itself. The region
has a tuple of arguments comprised of a function's external dependencies and its arguments, and a
tuple of results corresponding to a function's results.

An \applynode represents a function invocation. Its first input takes a \lambdanode's output
as origin, and all other inputs represent the function arguments. In the rest of the paper, we
refer to an \applynode's first input as its \emph{function input}, and to all its other inputs as
its \emph{argument inputs}. Invocation maps the values of a $\lambda$-node's input $k$-tuple to the
first $k$ arguments of the \lambdaregion, and the values of the function arguments of the
\applynode to the rest of the arguments of the \lambdaregion. The function body is evaluated and
the values of the \lambdaregion's results are mapped to the outputs of the \applynode.

\begin{figure*}
	\subfloat[RVSDG with \lambdam and \deltanodes\label{fig:lambda}]{
		\begin{minipage}{0.5\textwidth}
			\centering
			\includegraphics[scale=0.82, valign=c]{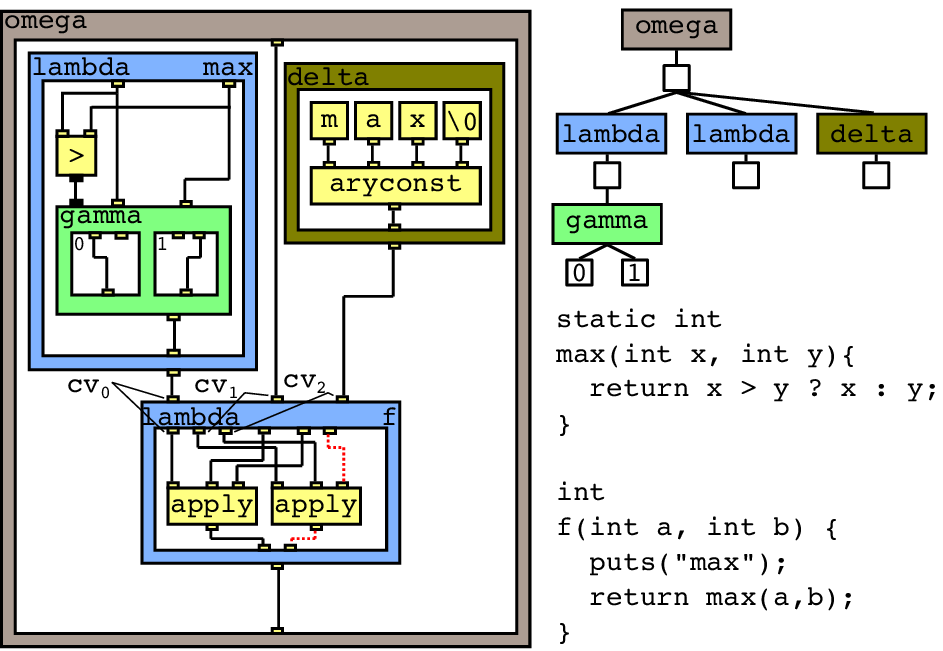}
			\vphantom{\includegraphics[scale=0.82, valign=c]{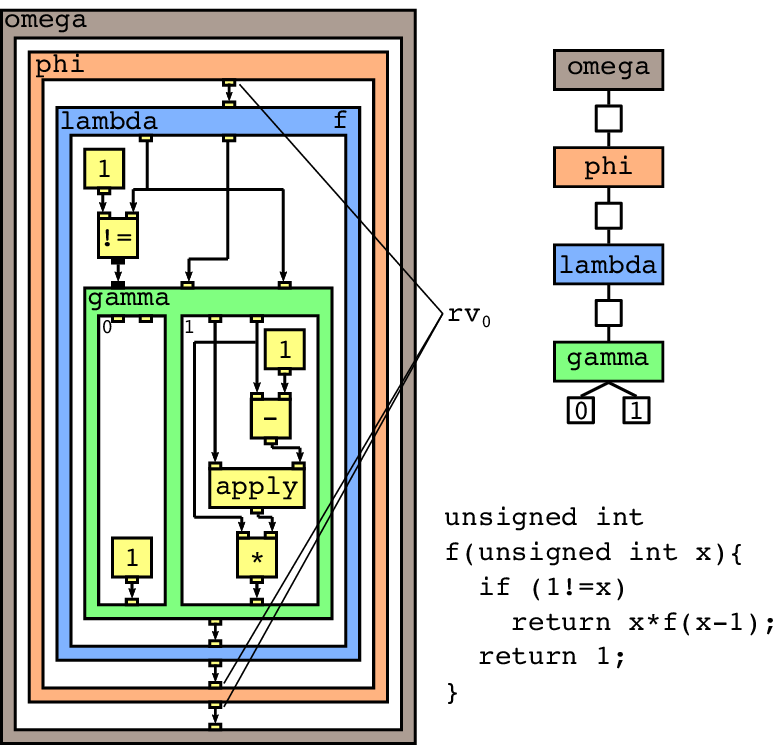}}
		\end{minipage}
	}
	\subfloat[RVSDG with a \phinode\label{fig:phi}]{
		\begin{minipage}{0.5\textwidth}
			\centering
			\includegraphics[scale=0.82, valign=c]{phi}
		\end{minipage}
	}
	\caption{Example for the usage of \lambdam, \deltam, and \phinodes, as well as corresponding
		region trees.}
	\label{fig:example2}
\end{figure*}

Figure \ref{fig:lambda} shows an RVSDG with two \lambdanodes. Function $f$ calls functions $puts$
and $max$ with the help of \applynodes. The function $max$ is part of the translation unit, while
$puts$ is external and must be imported (see the paragraph about \omeganodes for more details).
We further define the \emph{context variable} of a \lambdanode. A context variable provides the
corresponding input and argument for a variable a \lambdanode depends on.

\begin{definition}
The pair $cv_{l} = (\iput{}{l}, \argument{}{l})$ is a \lambdanode's $l$-th \emph{context variable}.
It consists of the $l$-th input and argument. We refer to the set of all context
variables as $CV$.
\end{definition}

Figure~\ref{fig:lambda} shows the three context variables of fuction $f$ annotated: one for
function $max$, one for function $puts$, and one for the global variable representing the string
argument to $puts$.

\begin{definition}
The \lambdanode connected to a function input is the \emph{callee} of an \applynode, and an
\applynode is the \emph{caller} of a \lambdanode. We refer to the set of all callers of a
\lambdanode as $CLL$.
\end{definition}

\subsubsection{Delta-Nodes}
A \deltanode models a global variable and contains a single region representing the constants'
value. It features a tuple of inputs and a single output. The inputs refer to the external
variables the \deltanode depends on, and the output represents the \deltanode itself. The region
has a tuple of arguments representing a global variable's external dependencies and a single
result corresponding to its right-hand side value.

Figure~\ref{fig:lambda} shows an RVSGD with a \deltanode. Function $puts$ takes a string as
argument that is the right-hand side of a global variable.
Similarly to \lambdanodes, we define the \emph{context variable} of a \deltanode. It provides the
corresponding input and argument for a variable a \deltanode depends on.

\begin{definition}
The pair $cv_{l} = (\iput{}{l}, \argument{}{l})$ is a \deltanode's $l$-th \emph{context variable}.
It consists of the $l$-th input and argument. We refer to the set of all context variables as $CV$.
\end{definition}

\subsubsection{Phi-Nodes}
A \phinode models an environment with mutually recursive functions, and contains
a single region with \lambdanodes. Each single output of these \lambdanodes serves as origin to a
single result in the \phiregion. A \phinode's outputs expose the individual functions to callers
outside the \phiregion, and must therefore have the same arity and signature as the results of the
\phiregion. The first input of an \applynode from \emph{outside} the \phiregion takes these
outputs as origin to invoke one of the functions.

The inputs of a \phinode refer to variables that the contained functions depend on and are mapped
to corresponding arguments in the \phiregion when a function is invoked. In addition, a
\phiregion has arguments for each contained function. An \applynode from
inside a \phiregion takes these as origin to its function input.

\phinodes permit a program's mutually recursive functions to be expressed in the RVSDG
without the introduction of cycles. Figure \ref{fig:phi} shows an RVSDG with a \phinode.
The function $f$ calls itself, and therefore needs to be in a \phinode to preserve the RVSDG's
acyclicity. The region in the \phinode has one input, representing the declaration of $f$, and
one output, representing the definition of $f$. The \phinode has one output so that $f$ can
be called from outside the recursive environment.

We define \emph{context variables} and \emph{recursion variables}. Context variables
provide corresponding inputs and arguments for variables the \lambdanodes from within a \phiregion
depend on. Recursion variables provide the argument and output an \applynode's function
input connects to.

\begin{definition}
The pair $cv_{l} = (\iput{}{l}, \argument{}{l})$ is the $l$-th \emph{context variable} of a
\phinode. It consists of the $l$-th input and argument. We call the set of all context variables
$CV$.
\end{definition}

\begin{definition}
For a \phinode with $n$ context variables, the triple $rv_{l} = (\result{}{l}, \argument{}{l+n},
\oput{}{l})$ is the $l$-th \emph{recursion variable}. It consists of the $l$-th result and
$l+n$-th argument of the \phiregion as well as the $l$-th output of the \phinode. We refer to the
set of all recursion variables as $RV$.
\end{definition}

Figure~\ref{fig:phi} shows the the \phinode's recursion variable annotated.

\subsubsection{Omega-Nodes}
An \omeganode models a translation unit. It is the top-level node of an
RVSDG and has no inputs or outputs. It contains exactly one region. This region's
arguments represent entities that are external to the translation unit and therefore need to be
imported. Its results mark all exported entities in the translation unit. Figure
\ref{fig:lambda} and \ref{fig:phi} illustrate the usage of \omeganodes. The \omegaregion in Figure
\ref{fig:lambda} has one argument, representing the import of function \verb;g;, and one result,
representing the export of function \verb;f;. The \omegaregion in Figure \ref{fig:phi} has only
one export for function \verb;f;.

\subsection{Edges}\label{sec:edges}
Edges connect node outputs or region arguments to a node input or region result, and are either
value typed, \ie, represent the flow of data between computations, or state typed, \ie,
impose an ordering on operations with side-effects. State edges are used to preserve
the observational semantics of the input program by ordering its side-effecting operations. Such
operations include memory read and writes, as well as exceptions.

In practice, a richer type system permits further distinction between different kind of values or
states. For example, different types for fixed- and floating-point values helps to distinguish
between these arithmetics, and a type for functions permits to correctly specify the
output types of \lambdanodes and the function input of \applynodes.

\section{Construction \& Destruction}

RVSDG construction and destruction generate an RVSDG from an input program
and reestablish control flow for code generation, respectively. We present both stages with an
\emph{Inter-Procedure Graph} (IPG) and a CFG as input and output. The IPG is an extension of a
call graph and captures all static dependencies between functions and global variables,
incorporating not only those originating from (direct) calls, but also those from other references
within a function. In the IPG, an edge from node $n1$ to node $n2$ exists, if the body of a
function/global variable corresponding to $n1$ references a function/global variable represented
by $n2$. The utilization of an IPG and a CFG permits a language-independent presentation of RVSDG
construction and destruction.

\subsection{Construction}\label{sec:construction}
RVSDG construction maps all constructs, concepts, and abstractions of an
input language to the RVSDG. The mapping is language-specific, and depends on the language's
features. Languages that permit unstructured control flow, such as C
or C++, cannot be mapped directly to the RVSDG and require a CFG as a stepping
stone, while languages such as Haskell permit direct construction~\cite{Reissmann12}.
In this section, we present RVSDG construction for the former case as it
encompasses the latter.
Conceptually, RVSDG construction can be split in two phases:

\begin{enumerate}
	\item \textit{Inter-Procedural Translation} (Inter-PT) translates functions, global variables,
	and inter-procedural dependencies, creating \lambdam, \deltam, and \phinodes.

	\item \textit{Intra-Procedural Translation} (Intra-PT) translates intra-procedural
		control and data flow, creating a \deltam/\lambdaregion from a function's/global variables'
		body.
\end{enumerate}

Inter-PT invokes Intra-PT for each function's or global variables' body, and both
phases interact through a common symbol table. The table maps function and CFG variables to
RVSDG arguments or outputs, and is updated with every creation of a
node or region. We omit the updates from our algorithm descriptions for brevity.

\subsubsection{Inter-Procedural Translation}\label{sec:interproc}

\begin{figure*}
	\subfloat[Code\label{fig:exp-ipg-code}] {
		\includegraphics[scale=0.84]{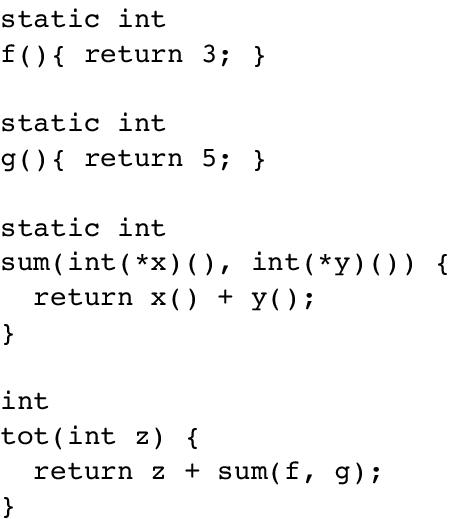}
	}
	\hfill
	\subfloat[IPG\label{fig:exp-ipg-ipg}] {
		\includegraphics[scale=0.84]{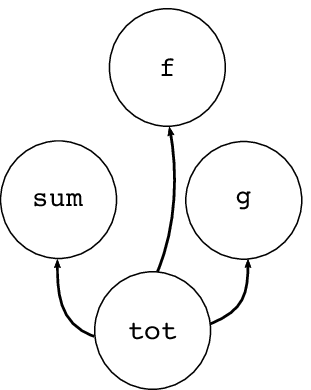}
	}
	\hfill
	\subfloat[RVSDG\label{fig:exp-ipg-rvsdg}] {
		\includegraphics[scale=0.84]{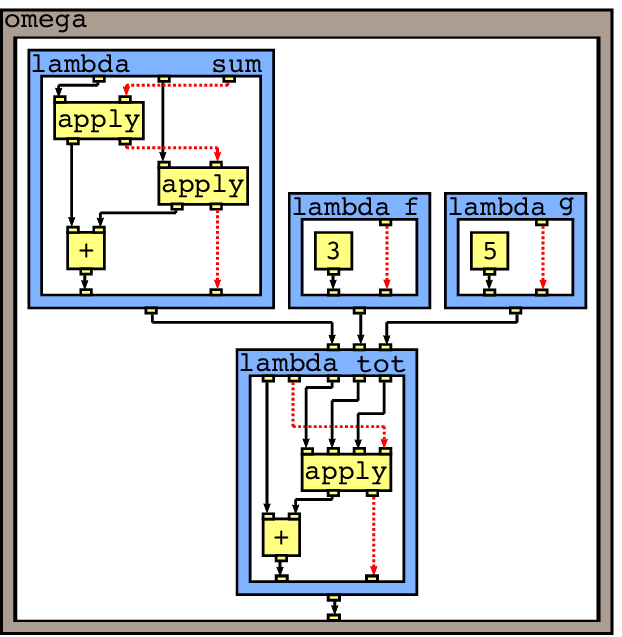}
	}
	\caption{Inter-Procedural Translation}
	\label{fig:exp-ipg}
\end{figure*}

Inter-PT converts all functions and global variables from the \emph{Inter-Procedure Graph} (IPG) of
a translation unit to \lambdanodes and \deltanodes, respectively. Figure~\ref{fig:exp-ipg-ipg}
shows the IPG for the code in Figure~\ref{fig:exp-ipg-code}. The code
consists of four functions, with function \verb;sum; performing two indirect calls. The
corresponding IPG consists of four nodes and three edges. All edges originate from node $tot$, as
it is the only function that explicitly references other functions, \ie \verb;sum; for a direct
call, and \verb;f; and \verb;g; to pass as argument. No edge originates from node \verb;sum;, as
the corresponding function does not explicitly reference any other functions, and the functions for
the indirect calls are provided as arguments.

The RVSDG puts two constraints on the translation from an IPG. Firstly, mutually recursive entities
are required to be created within \phinodes to preserve the RVSDG's acyclicity. Secondly, Inter-PT
must respect the calling dependencies of functions to ensure that \lambdanodes are created before
their \applynodes. In order to embed mutually recursive entities into \phinodes, we need to
identify the strongly connected components (SCCs) in the IPG. We consider an SCC
\emph{trivial}, if it consists only of a single node with no self-referencing edges. Otherwise, it
is \emph{non-trivial}. Moreover, a trivial SCC might not have a CFG associated with it, and is
therefore defined in another translation unit.

\begin{informalg}{Inter-Procedural Translation}{alg:interpt}
Compute all SCCs in an IPG and process them in topological order of the directed acyclic
graph formed by the SCCs as follows:
	\begin{enumerate}
		\item \label{item:trivial} \bpoint{Trivial SCC}:
			\begin{enumerate}
				\item \bpoint{Function with CFG}: Begin a \lambdanode by adding all context variables,
					function arguments, and an additional state argument to the \lambdaregion. Translate the
					CFG with Intra-PT as explained in Section~\ref{sec:intraproc}, and finish the
					\lambdanode by adding the function results and the state result to the
					\lambdaregion. If a function is exported, add a result to the \omegaregion and connect
					the \lambdanode's output to it.

				\item \bpoint{Global variable with CFG}: Begin a \deltanode by adding all context variables
					to the \deltaregion. Translate the CFG with Intra-PT as eplained in
					Section~\ref{sec:intraproc}, and finish the \deltanode by adding the result to the
					\deltaregion. If a global variable is exported, add a result to the \omegaregion and
					connect the \deltanode's output to it.

				\item \bpoint{Without CFG}: Add a \omegaregion argument for the external entity.
			\end{enumerate}

		\item \label{item:non-trivial} \bpoint{Non-trivial SCC}: Begin a \phinode by adding all
			functions/global variables as well as context variables to the \phiregion. Translate each
			entity in the SCC according to \textsc{Trivial SCC} without exporting them. Finish the
			\phinode by adding all outputs as results to the \phiregion. If an entity is exported, add a
			result to the \omegaregion and connect the \phinode's output to it.
		\vspace*{-\baselineskip}
	\end{enumerate}
\end{informalg}

Algorithm~\ref{alg:interpt} outlines the RVSDG construction from an IPG. It finds all SCCs
and converts trivial SCCs to individual \deltam/\lambdanodes, while the \deltam/\lambdanodes
created from non-trivial SCCs are embedded in \phinodes. This satisfies the first constraint. The
second
constraint is satisfied by processing SCCs in topological order, creating \lambdanodes before their
\applynodes. Identification and ordering of SCCs can be done in a single step with
Tarjan's algorithm~\cite{Tarjan72}, which returns identified SCCs in reverse topological order.
Figure~\ref{fig:exp-ipg-rvsdg} shows the RVSDG after application of Algorithm~\ref{alg:interpt}
to the IPG in Figure~\ref{fig:exp-ipg-ipg}. In addition to a function's arguments,
Algorithm~\ref{alg:interpt} adds a state argument and result to \lambdaregions (the
red dashed line in Figure~\ref{fig:exp-ipg-rvsdg}), to sequence stateful computations.
Nodes representing operations with
side-effects consume this state and produce a new state for the next node.

\subsubsection{Intra-Procedural Translation}\label{sec:intraproc}
The RVSDG puts several constraints on the translation of intra-procedural control and data flow.
Firstly, it requires that the control flow only consists of constructs that can be translated to
\gammam and \thetanodes, \ie it can only consist of tail-controlled loops and conditionals with
symmetric control flow splits and joins. Secondly, the nesting and relation of these
constructs to each other is required as the RVSDG is a hierarchical representation. Thirdly, it is
necessary to know the data dependencies of these structures in order to construct \gammam and
\thetanodes. While these constraints are beneficial for optimizations by substantially simplifying
their implementation%
, they render RVSDG construction non-trivial.

This section's construction algorithm enables the translation of any data and control flow,
irregardless of its complexity, to the RVSDG. It creates a \deltam/\lambdaregion from a global
variables' or function's body in four stages:

\begin{enumerate}
	\item \textit{Control Flow Restructuring} (CFR) restructures a CFG to make
		it amenable to RVSDG construction.

	\item \textit{Structural Analysis} constructs a control tree~\cite{Muchnick97} from the
		restructured CFG, discovering the CFG's individual control flow regions.

	\item \textit{Demand Annotation} annotates the discovered control flow regions with the
		variables that are demanded by the instructions within these regions.

	\item \textit{Control Tree Translation} converts the annotated control tree into a
		\deltam/\lambdaregion.
\end{enumerate}

CFR ensures the first requirement by translating a function's control flow to a form that is
amenable to RVSDG construction. It restructures control flow to a form that enables the direct
mapping of a CFG's control flow regions to the RVSDG's \gammam and \thetanodes. CFR can be omitted
for languages with limited control flow structures, such as Haskell or Scheme. Structural analysis
ensures the second requirement by constructing a control tree from the CFG, exposing the control
regions nesting and the relation to each other. Demand annotation fulfills the third requirement
by annotating the control tree's nodes with their data dependencies. Finally, the annotated control
tree can be translated to a \deltam/\lambdaregion. The rest of this section covers the four stages
in detail.

\paragraph{Control Flow Restructuring:}
CFR converts a CFG to a form that only contains tail-controlled loops and conditionals with
properly nested splits and joins. This stage is only necessary for languages that support more
complex control flow constructs, such as \verb;goto; statements or short-circuit operators, but
can be omitted for languages with more limited control flow. CFR consists of two interlocked
phases: loop restructuring and branch restructuring. Loop restructuring transforms all loops to
tail-controlled loops, while branch restructuring ensures conditionals with symmetric control flow
splits and joins.

We omit an extensive discussion of CFR as it is detailed in Bahmann~\etal~\cite{Bahmann15}.
In contrast to node splitting approaches~\cite{Zhang04}, CFR avoids the possibility of exponential
code blowup~\cite{Carter03} by inserting additional predicates and branches instead of cloning
nodes. Moreover, it does not require a CFG in SSA form as this form is automatically established
throughout construction.

\paragraph{Structural Analysis:}
After CFR, a restructured CFG consists of 3 single-entry/single-exit control flow regions:

\begin{itemize}
	\item \textit{Linear Region}: A linear subgraph where the entry node and all intermediate nodes
		have only one outgoing edge, and the exit node as well as all intermediate nodes have only one
		incoming edge.

	\item \textit{Branch Region}: An subgraph with the entry and exit node representing the
		control flow split and join, respectively, and each branch alternative consisting of a single
		node.

	\item \textit{Loop Region}: A single node where an edge originates and targets this node.
\end{itemize}

These control flow regions and their corresponding nesting structure can be exposed by performing
an interval~\cite{Muchnick97} or structural~\cite{Sharir80} analysis. The analysis result is a
control tree~\cite{Muchnick97} with basic blocks as leaves and abstract nodes representing the
control flow regions as branches.

A linear region maps to a \emph{linear node} in the control tree with the
linear subgraph's entry and exit node as the node's left and right most child, respectively.
A branch region maps to two control tree nodes: a \emph{branch node} and a linear node. The branch
node represents the region's alternatives with the corresponding nodes as its children.
A linear node with three children can then be used to capture the rest of the branch region. Its
first child is the region's entry node, the second child the branch node representing the
alternatives, and the third child the region's exit node. Finally, a loop region maps to a
\emph{loop node} with the region's single node as its child.

\begin{figure*}
	\subfloat[CFG\label{fig:euclid_cfg}]{
		\includegraphics[scale=0.84]{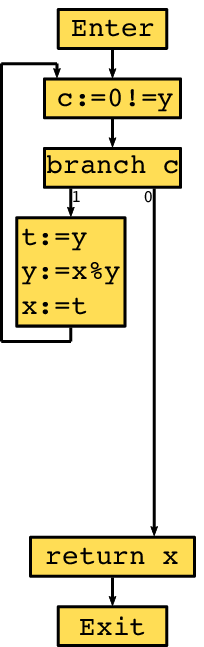}
	}
	\hfill
	\subfloat[Restructured CFG\label{fig:euclid_restructured}]{
		\includegraphics[scale=0.84]{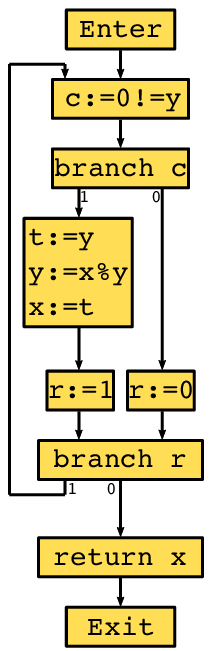}
	}
	\hfill
	\subfloat[Annotated control tree\label{fig:euclid_pst}]{
		\includegraphics[scale=0.84]{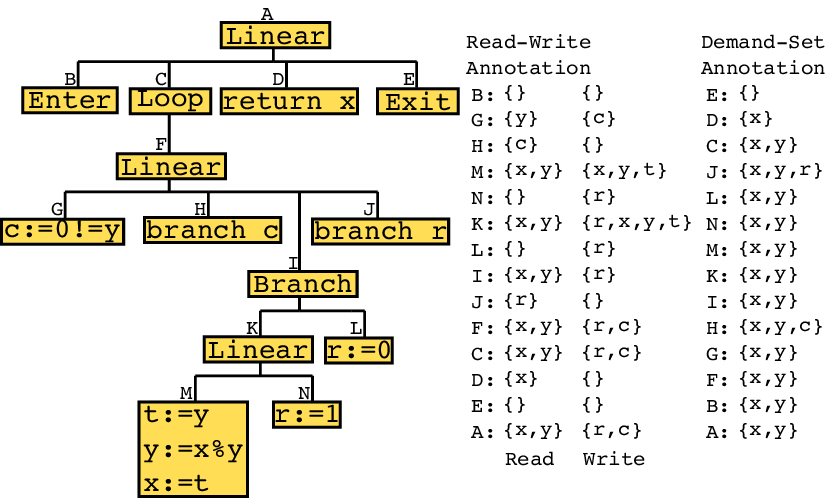}
	}
	\hfill
	\subfloat[RVSDG\label{fig:euclid_rvsdg}]{
		\includegraphics[scale=0.84]{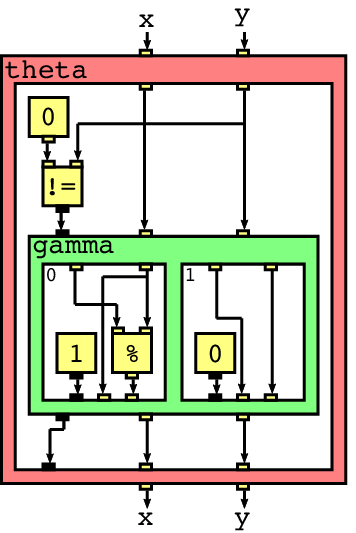}
	}
	\caption{Intra-Procedural Translation}
	\label{fig:euclid}
\end{figure*}

Figure~\ref{fig:euclid_cfg} shows Euclid's algorithm as a CFG, and
Figure~\ref{fig:euclid_restructured} shows the same CFG after CFR, which restructured the
head-controlled loop to a tail-controlled loop. The left of Figure~\ref{fig:euclid_pst} shows
the corresponding control tree. %

\paragraph{Demand Annotation:}
Structural analysis exposes necessary control flow regions for a direct translation to
RVSDG. A control flow tree's branch and loop nodes map directly to \gammam and
\thetanodes, and individual instructions to simple nodes, but it is further
necessary to expose the data dependencies of these nodes for efficient generation.

This is the task of demand annotation. It exposes these data dependencies by annotating control
tree nodes with the variables that are demanded by the instructions within control flow regions.
It accomplishes this using a \emph{read-write} and \emph{demand-set} annotation pass.
The read-write pass annotates each control tree node with the set of read and written variables of
the corresponding control flow region, while the demand-set pass uses these variables
to annotate each control tree node with the set of demanded variables, \ie variables that are
necessary to fulfill the dependencies of the instructions within a control flow region.

\begin{informalg}{Demand Annotation}{alg:da}
\begin{enumerate}
	\vspace*{-0.7\baselineskip}

	\item \bpoint{Read-Write Annotation}: Process the control tree nodes in post-order as follows:
		\begin{itemize}
			\item \bpoint{Basic Block}: For each instruction $i$ processed bottom-up, the read set is
				$R = (R \setminus W_i) \cup R_i$. The write set is $W = \bigcup W_i$.

			\item \bpoint{Linear Node}: For each child $c$ processed right to left, the read set is
				$R = (R \setminus W_c) \cup R_c$. The write set is $W = \bigcup W_c$.

			\item \bpoint{Branch Node}: For each child $c$, the read set and write set is
				$R = \bigcup R_c$ and $W = \bigcap W_c$, respectively.

			\item \bpoint{Loop Node}: For the child $c$, the read set and write set is $R = R_c$ and
				$W = W_c$, respectively.
		\end{itemize}

	\item \bpoint{Demand-Set Annotation}: Process the control tree nodes with an empty demand set
		$D_t$ as follows:
		\begin{itemize}
			\item \bpoint{Basic Block}: Set $D = D_t = (D_t \setminus W) \cup R$ and continue processing.

			\item \bpoint{Linear Node}: Recursively process the children right to left.
				Set $D = D_t = (D_t \setminus W) \cup R$. and continue processing.

			\item \bpoint{Branch Node}: Set $D_{tmp} = D_t$. Recursively process each child with a copy
				of $D_t$. Set $D = D_t = (D_{tmp} \setminus W) \cup R$ and continue processing.

			\item \bpoint{Loop Node}: Set $D = D_t \cup R$. Recursively process the child with
				$D_t = D$ and continue processing.
		\end{itemize}

	\vspace*{-\baselineskip}
\end{enumerate}
\end{informalg}

Algorithm~\ref{alg:da} shows the details of the two passes. The read-write pass annotates each node
with the read set $R$ and write set $W$. It processes the tree in post-order, building up
the two sets from the innermost to the outermost nested control flow region. For linear nodes, the
children are processed from right to left, \ie bottom-up in the restructured CFG, to create the
two sets. For branch nodes, a variable is only considered to be written, if it is in the write
set of \emph{all} the node's children, \ie it was written in all alternatives of a conditional.

The demand-set pass uses the read set $R$ and write set $W$ to construct a demand set $D$ for each
node. The algorithm is initialized with an empty set $D_t$, which is used to keep track of
demanded variables during traversal. The demand-set pass traverses the tree such that it follows
a bottom-up traversal of the restructured CFG, adding and removing variables from $D_t$ during
this traversal according to each node's rules. For branch nodes, each child is processed with a copy of
$D_t$, as the corresponding alternatives of the conditional are independent from another. For loop
nodes, the \thetanode's requirement that inputs and outputs must have the same signature
necessitates that $R$ is added to $D_t$ \emph{before} the loop's body is processed.
The right of Figure~\ref{fig:euclid_pst} shows the traversal order for the two passes along with
the read, write, and demand set for each node of the control tree on the left.

\paragraph{Control Tree Translation:}
After demand annotation, each node of the control tree is annotated with the set of variables that
its instructions require, \ie their data dependencies. Finally, the control tree translation
constructs a \deltam/\lambdaregion from the control tree along with its annotated demand sets.
Algorithm~\ref{alg:st} shows the details. The algorithm processes each node in the control tree
creating \gammam and \thetanodes for all branch and loop nodes, respectively.
It uses the demand set of the right sibling for the outputs of gamma nodes,
corresponding to the branch region's join node in the CFG.
Figure~\ref{fig:euclid_rvsdg} shows the resulting RVSDG nodes for the example.

\begin{informalg}{Control Tree Translation}{alg:st}
Process the control tree nodes as follows:
\begin{itemize}
	\item \bpoint{Basic Block}: Process the node's operations top-down creating simple nodes in the
		RVSDG.

	\item \bpoint{Linear Node}: Recursively process the node's children top-down.

	\item \bpoint{Branch Node}:Begin a \gammanode with inputs according to the node's demand set.
		Create subregions by recursively processing the node's children. Finish the \gammanode with
		outputs according to its right sibling node's demand set.

	\item \bpoint{Loop Node}: Begin a \thetanode with inputs according to the node's demand set.
		Create its region by recursively processing its child. Finish the \thetanode with outputs
		according to its demand set.

	\vspace*{-\baselineskip}
\end{itemize}
\end{informalg}

\subsubsection{Modeling Stateful Computations}\label{sec:modeling-state}
Algorithm~\ref{alg:interpt} adds an additional state argument and result to every \lambdanode. This
state is used to sequentialize all stateful computations within a function. Nodes
with side-effects consume this state and produce a new state for consumption by the next node. This
single state ensures that the order of operations with side-effects in the RVSDG is according to
the total order specified in the original program, ensuring correct observable behavior.
Specifically, the use of a single state for sequentializing stateful operations ensures that the
order of these operations in the RVSDG is equivalent to the order in the restructured CFG.

The utilization of a single state is, however, overly
conservative, as different computations can have mutually exclusive side-effects. For example, the
side-effect of a non-terminating loop is unrelated to a non-dereferencable load. These stateful
computations can be modeled independently with the help of distinct states, as depicted in
Figure~\ref{fig:exp2-rvsdg}. This results in the explicit exposure of more concurrent computations,
as loops with no memory operations would become independent from other loops with memory
operations. Moreover, the possibility of encoding independent states can also be leveraged by
analyses and optimizations. For example, alias analysis can directly encode independent memory
operations into the RVSDG by introducing additional memory states. Pure functions could
be easily recognized and optimized, as they would contain no operations that use the added states
and therefore would only pass it through, \ie, the origin of the state result would be the
\lambdaregion's argument.

\subsection{Destruction}\label{sec:destruction}
The destruction stage reestablishes control flow by extracting an IPG from an RVSDG
and generating CFGs from individual \lambdaregions. Inter-Procedural Control Flow Recovery
(Inter-PCFR) creates an IPG from \lambdanodes, while
Intra-Procedural Control Flow Recovery (Intra-PCFR) extracts control flow from \gammam and
\thetanodes and generates basic blocks with corresponding operations for primitive nodes.
A \lambdaregion without \gammam and \thetanodes trivially transforms into a linear CFG,
while \lambdaregions with these nodes require construction of branches and/or loops.
This section discusses Inter-PCFR in detail. Detailed discussion
of Intra-PCFR is found in Bahmann~\etal~\cite{Bahmann15}.

\subsubsection{Inter-Procedural Control Flow Recovery}\label{sec:ipcfr}
Inter-PCFR recovers an IPG from an RVSDG. IPG nodes are created for \lambdanodes as well as
arguments of the \omegaregion, while IPG edges are inserted to capture the dependencies between
\lambdanodes. Algorithm~\ref{alg:intrapt} starts by
creating IPG nodes for all arguments of the \omegaregion, \ie, all external functions. It
continues by recursively traversing the region tree, creating IPG nodes for encountered
\lambdanodes and IPG edges for their dependencies. For the region of every \lambdanode, it
invokes Intra-PCFR to create a CFG.

\begin{informalg}{Inter-Procedural Control Flow Recovery}{alg:intrapt}
\begin{enumerate}
	\vspace*{-0.7\baselineskip}

	\item Create IPG nodes for all arguments of the \omegaregion.
	\item Process all nodes of the \omegaregion in topological order as follows:
		\begin{itemize}
			\item \bpoint{\lambdanodes}: Create an IPG node, and mark it exported if the
				\lambdanode's output has a \omegaregion's result as user. For every context variable
				$cv = (\iput{}{}, \argument{}{})$, add an edge from the \lambdanode's IPG node to the
				corresponding IPG node of the producer of $\iput{}{}$. Create a CFG from the \lambdanode's
				subregion and attach it to the IPG node.

			\item \bpoint{\deltanodes}: Create an IPG node, and mark it exported if the
				\deltanode's output has a \omegaregion's result as user. For every context variable
				$cv = (\iput{}{}, \argument{}{})$, add an edge from the \deltanode's IPG node to the
				corresponding IPG node of the producer of $\iput{}{}$. Create the expression from the
				\deltanode's subregion and attach it to the IPG node.

			\item \bpoint{\phinodes}: For every argument of the \phiregion, create an IPG node for
				the corresponding \deltam/\lambdanode and add IPG edges from this node to the
				corresponding IPG nodes of the context variables. Translate the \deltam/\lambdanodes in the
				\phiregion according to the rules above. Mark the IPG node as exported if the corresponding
				\phinode's output has a \omegaregion's result as user.
		\end{itemize}

	\vspace*{-\baselineskip}
\end{enumerate}
\end{informalg}

\subsubsection{Intra-Procedural Control Flow Recovery}
Bahmann~\etal~\cite{Bahmann15} explored two different approaches for CFG generation:
\textit{Structured Control Flow Recovery (SCFR)} and \textit{Predicative Control Flow Recovery
(PCFR)}. SCFR uses the region hierarchy within a \lambdaregion to recover control flow, while PCFR
generates branches for predicate producers and follows the predicate consumers to the eventual
destination. Both schemes reestablish evaluation-equivalent CFGs, but differ in the recoverable
control flow. SCFR recovers only control flow that resembles the structural nodes in
\lambdaregions, \ie, control flow equivalent to \verb;if-then-else;, \verb;switch;, and
\verb;do-while; statements, while PCFR can recover arbitrary complex control flow, \ie, control
flow that is not restricted to RVSDG constructs. PCFR reduces the number of static branches in the
resulting control flow~\cite{Bahmann15}, but might also result in undesirable
control flow for certain architectures, such as graphic processing units~\cite{Reissmann16}.
For the sake of brevity, we omit a discussion of SCFR and PCFR as the algorithms are extensively
described by Bahmann~\etal~\cite{Bahmann15}.

\section{Optimizations}\label{sec:optimizations}
The properties of the RVSDG make it an appealing IR for optimizing compilers. Many optimizations
can be expressed as simple graph traversals, where subgraphs are rewritten, nodes are moved between
regions, nodes or edges are marked, or edges are diverted. In this section, we present Common and
Dead Node Elimination optimizations that exploit the RVSDG's properties to unify traditionally
distinct transformations.

\subsection{Common Node Elimination}\label{sec:cne}
Common Node Elimination (CNE) permits the removal of redundant computations by detecting congruent
nodes. These nodes always produce the same results, enabling the redirection of their result edges
to a single node. This renders the other nodes dead, permitting Dead Node Elimination to remove
them. CNE is similar
to common subexpression elimination and value numbering~\cite{Alpern88} in that it detects
equivalent computations, but as the RVSDG represents all computations uniformly as nodes, it can
be extended to conditionals~\cite{Rugina00}, loops, and functions.

We consider two simple nodes $n_1$ and $n_2$ congruent, or $n_1 \cong n_2$, if they represent the
same computation, have the same number of inputs, \ie, $\card{I_{n_1}} = \card{I_{n_2}}$, and the
inputs $\iput{k}{n_1}$ and $\iput{k}{n_2}$ are congruent, or $\iput{k}{n_1} \cong \iput{k}{n_2}$,
for all $k = [0..\card{I_{n_1}}]$. Two inputs are congruent if their respective origins
$\origin{k}{n_1}$ and $\origin{k}{n_2}$ are congruent, \ie, $\origin{k}{n_1} \cong
\origin{k}{n_2}$. By definition, the origins of inputs are either outputs of simple or structural
nodes, or arguments of regions. Origins from simple nodes are only equivalent when their respective
producers are computationally equivalent, whereas for the other cases, it must be guaranteed that
they always receive the same value.

The implementation of CNE consists of two phases: mark and divert. The mark phase identifies
congruent simple nodes, while the divert phase diverts all edges of their origins to
a single node, rendering all other nodes dead. Both phases of Algorithm~\ref{alg:cne} perform a
simple top-down traversal, recursively processing subregions of structural nodes
annotating inputs, outputs, arguments, and results, as well as simple nodes as congruent. For
\gammanodes, the algorithm marks only computations within a \emph{single} region as congruent and
performs no analysis between regions. In the case of \thetanodes, computations are only congruent
when they are congruent before and after the loop execution, \ie, the inputs and results of two
loop variables must be congruent.
Figure~\ref{fig:dcne1} shows the RVSDG for the code in Figure~\ref{fig:dcne-code}, and Figure
\ref{fig:dcne2} the RVSDG after CNE. Two of the four multiplications take the same inputs and
therefore are congruent to each other, resulting in the redirection of their result edges.

\begin{informalg}{Common Node Elimination}{alg:cne}
\begin{enumerate}
	\vspace*{-0.7\baselineskip}

	\item \bpoint{Mark}: Process all nodes in topological order as follows:
		\begin{itemize}
			\item \bpoint{Simple nodes}: Denote this node as \n. Mark \n as congruent to all nodes
				\np which represent the same operation and where $\card{I_n} = \card{I_{n^{\prime}}} \land
				\iput{k}{n} \cong \iput{k}{n^{\prime}}$ for all $k = [0..\card{I_n}]$. Mark all outputs
				$\oput{k}{n} \cong \oput{k}{n^{\prime}}$ for all $k = [0..\card{O_n}]$.

			\item \bpoint{\gammanode}: For all entry variables $ev_1, ev_2 \in EV$ where
				$\iput{}{ev_1} \cong \iput{}{ev_2}$, mark $\argument{k}{ev_1} \cong \argument{k}{ev_2}$ for
				all $k \in [0..\card{A_{ev_1}}]$. Recursively process the \gammaregions. For all exit
				variables $ex_1, ex_2 \in EX$ where $\result{k}{ex_1} \cong \result{k}{ex_2}$ for all
				$k \in [0..\card{R_{ex_1}}]$, mark $\oput{}{ex_1} \cong \oput{}{ex_2}$.

			\item \bpoint{\thetanode}: For all loop variables $lv_1, lv_2 \in LV$ where
				$\iput{}{lv_1} \cong \iput{}{lv_2} \land \result{}{lv_1} \cong \result{}{lv_2}$, mark
				$\argument{}{lv_1} \cong \argument{}{lv_2}$ and $\oput{}{lv_1} \cong \oput{}{lv_2}$.
				Recursively process the \thetaregion.

			\item \bpoint{\lambdanode}: For all context variables $cv_1, cv_2 \in CV$ where
				$\iput{}{cv_1} \cong \iput{}{cv_2}$, mark $\argument{}{cv_1} \cong \argument{}{cv_2}$.
				Recursively process the \lambdaregion.

			\item \bpoint{\phinode}: For all context variables $cv_1, cv_2 \in CV$ where
				$\iput{}{cv_1} \cong \iput{}{cv_2}$, mark $a_{cv_1} \cong a_{cv_2}$. Recursively process
				the \phiregion.

			\item \bpoint{\omeganode}: Recursively process the \omegaregion.
		\end{itemize}

	\item \bpoint{Divert}: Process all nodes in topological order as follows:
		\begin{itemize}
			\item \bpoint{Simple nodes}: Denote this node as \n. For all nodes \np which are congruent to
				\n, divert all outputs $\oput{k}{n^{\prime}}$ to \oput{k}{n} for all $k = [0..\card{O_n}]$.

			\item \bpoint{\gammanode}: For all entry variables $ev_1, ev_2 \in EV$ where
				$\iput{}{ev_1} \cong \iput{}{ev_2}$, divert all edges from $\argument{k}{ev_2}$ to
				$\argument{k}{ev_1}$ for all $k \in [0..\card{A_{ev_1}}]$. Recursively
				process the \gammaregions. For all exit variables $ex_1, ex_2 \in EX$ where
				$\result{k}{ex_1} \cong \result{k}{ex_2}$ for all $k \in [0..\card{R_{ex_1}}]$, divert all
				edges from $\oput{}{ex_2}$ to $\oput{}{ex_1}$.

			\item \bpoint{\thetanode}: For all induction variables $lv_1, lv_2 \in LV$ where
				$\argument{}{lv_1} \cong \argument{}{lv_2} \land \oput{}{lv_1} \cong \oput{}{lv_2}$, divert
				all edges from $\argument{}{lv_2}$ to $\argument{}{lv_1}$ and from $\oput{}{lv_2}$ to
				$\oput{}{lv_1}$. Recursively process the \thetaregion.

			\item \bpoint{\lambdanode}: For all context variables $cv_1, cv_2 \in CV$ where
				$\iput{}{cv_1} \cong \iput{}{cv_2}$, divert all edges from $\argument{}{cv_2}$ to
				$\argument{}{cv_1}$. Recursively process the \lambdaregion.

			\item \bpoint{\phinode}: For all context variables $cv_1, cv_2 \in CV$ where
				$\iput{}{cv_1} \cong \iput{}{cv_2}$, divert all edges from $\argument{}{cv_2}$ to
				$\argument{}{cv_1}$. Recursively process the \phiregion.

			\item \bpoint{\omeganode}: Recursively process the \omegaregion.
		\end{itemize}

	\vspace*{-\baselineskip}
\end{enumerate}
\end{informalg}

For simple nodes, the algorithm marks all nodes within a
region that are congruent to a node \n. In order to avoid costly traversals of all nodes for every
node \n, the mark phase takes the candidates from the users of the origin of \n's first
input. If there is another input from a simple node \np with the same operation and number of
inputs among them, the other inputs from both nodes can be compared for congruence. Moreover, a
region must store constant nodes, \ie nodes without inputs, separately from other nodes so that the
candidate nodes for constants are available. For commutative simple nodes, the inputs should be
sorted before their comparison.

The presented algorithm only detects congruent simple nodes within a region. For \gammanodes,
congruence can also exist between nodes of different \gammaregions, and extending the
algorithm would eliminate these redundancies. Another extension would
be to detect congruent structural nodes, to implement conditional fusion
\cite{Rugina00} and loop fusion~\cite{manjikian97}. In
the case of \gammanodes, it is sufficient to ensure that two nodes have congruent
predicates, while \thetanodes require congruence detection between
different \thetaregions to ensure that their predicates are the same.

\subsection{Dead Node Elimination}\label{sec:dne}

\begin{figure*}
	\subfloat[Code\label{fig:dcne-code}]{
		\includegraphics[scale=0.84]{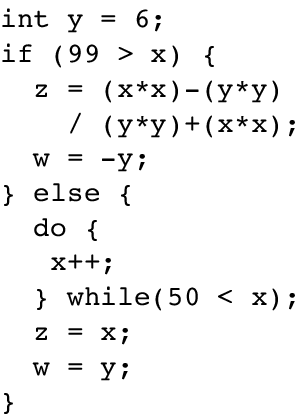}
	}
	\hfill
	\subfloat[RVSDG\label{fig:dcne1}]{
		\includegraphics[scale=0.84]{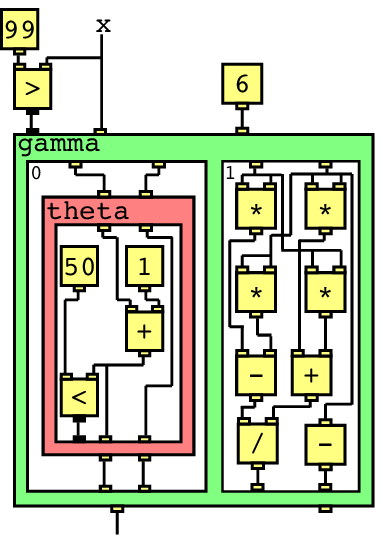}
	}
	\hfill
	\subfloat[After CNE\label{fig:dcne2}]{
		\includegraphics[scale=0.84]{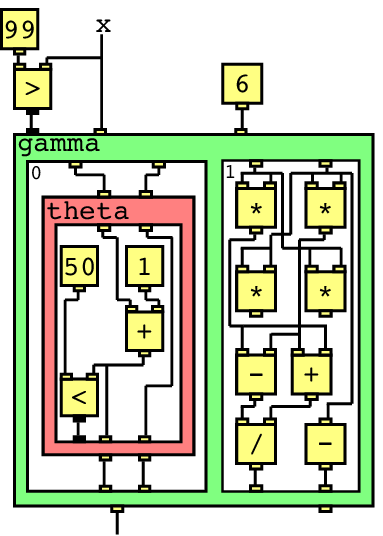}
	}
	\hfill
	\subfloat[After DNE mark\label{fig:dcne3}]{
		\includegraphics[scale=0.84]{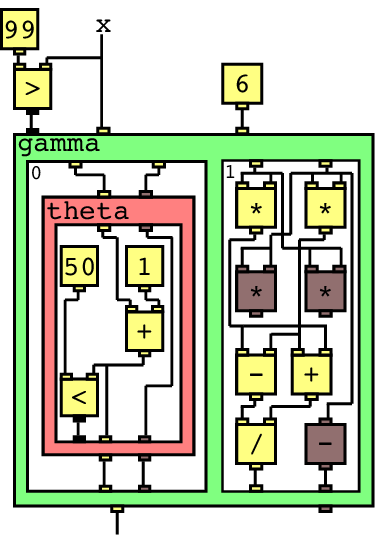}
	}
	\caption{Dead and Common Node Elimination}
	\label{fig:dcne}
\end{figure*}

Dead Node Elimination (DNE) is a combination of dead and unreachable code elimination, and removes
all nodes that do not contribute to the result of a computation. Dead nodes are generated by
unreachable and dead code from the input program, as well as by other
optimizations such as Common Node Elimination.
An operation is considered dead code when its results are either not used or only by
other dead operations. Thus, an output of a node is dead, if it has no users or all its users are
dead. We consider a node to be dead, if all its outputs are
dead. It follows that a node's inputs are dead, if the node itself is dead. We call all entities
that are not dead alive.

The implementation of DNE consists of two phases: mark and sweep. The mark phase identifies all
outputs and arguments that are alive, while the sweep phase removes all dead entities. The mark
phase traverses RVSDG edges according to the rules in Algorithm \ref{alg:dne}. If a structural
node is dead, the mark phase skips the traversal of its subregions as well as all of the contained
computations, as it never reaches the node in the first place. The mark phase is invoked for all
result origins of the \omegaregion.

The sweep phase performs a simple
bottom-up traversal of an RVSDG, recursively processing subregions of structural nodes as long as
these nodes are alive. A dead structural node is removed with all its contained computations.
The RVSDG's uniform representation of all computations as nodes permits DNE to not only remove
simple computations, but also compound computations such as conditionals, loops, or even entire
functions. Moreover, its nested structure avoids the processing of entire branches of the region
tree if they are dead.

\begin{informalg}{Dead Node Elimination}{alg:dne}
\begin{enumerate}
	\vspace*{-0.7\baselineskip}

	\item \bpoint{Mark}: Mark output or argument as alive and continue as follows:
		\begin{itemize}
			\item \bpoint{\omegaregion argument}: Stop marking.

			\item \bpoint{\phinode output}: Mark the result origin of the corresponding recursion
				variable.

			\item \bpoint{\phiregion argument}: Mark the input origin if the argument belongs to a
				context variable. Otherwise, mark the output of the corresponding recursion variable.

			\item \bpoint{\lambdanode output}: Mark all result origins of the \lambdaregion.

			\item \bpoint{\lambdaregion argument}: Mark the input origin if the argument is a dependency.

			\item \bpoint{\thetanode output}: Mark the \thetanode's predicate origin as well as the
				result and input origin of the corresponding loop variable.

			\item \bpoint{\thetaregion argument}: Mark the input origin and output of the corresponding
				loop variable.

			\item \bpoint{\gammanode output}: Mark the \gammanode's predicate origin as well as the
				origins of all results of the corresponding exit variable.

			\item \bpoint{\gammaregion argument}: Mark the input origin of the corresponding entry
				variable.

			\item \bpoint{Simple node output}: Mark the origin of all inputs.
		\end{itemize}

	\item \bpoint{Sweep}: Process all nodes in reverse topological order and remove them if they are
		dead. Otherwise, process them as follows:
		\begin{itemize}
			\item \bpoint{\omeganode}: Recursively process the \omegaregion. Remove all dead arguments.

			\item \bpoint{\gammanode}: For all exit variables $(R,o) \in EX$ where $o$ is dead, remove
				$o$ and all $r \in R$. Recursively process the \gammaregions. For all entry variables
				$(i,A) \in EV$ where all $a \in A$ are dead, remove all $a \in A$ and $i$.

			\item \bpoint{\thetanode}: For all loop variables $(i,a,r,o) \in LV$ where $a$ and $o$ are
				dead, remove $o$ and $r$. Recursively process the \thetaregion. Remove $i$ and $a$. 

			\item \bpoint{\lambdanode}: Recursively process the \lambdaregion. For all context variables
				$(i,a) \in CV$ where $a$ is dead, remove $a$ and $i$.

			\item \bpoint{\phinode}: For all recursion variables $(r, a, o) \in RV$ where $a$ and $o$ are
				dead, remove $o$ and $r$. Recursively process the \phiregion. Remove $a$. For all context
				variables $(i,a) \in CV$ where $a$ is dead, remove $a$ and $i$.
		\end{itemize}

	\vspace*{-\baselineskip}
\end{enumerate}
\end{informalg}

Figure~\ref{fig:dcne3} shows the RVSDG from Figure~\ref{fig:dcne2} after the mark phase.
Grey colored entities are dead. The mark phase traverses
the graph's edges, marking the \gammanode's leftmost output alive. This renders the
corresponding result origins of the \gammaregions alive, then the leftmost output of the
\thetanode, and so forth. After the mark phase annotated all outputs and arguments as alive, the
sweep phase removes all dead entities.

\section{Implementation and Evaluation}\label{sec:eval}
This section aims to demonstrate that the RVSDG has no inherent
impediment that prevents it from producing competitive code, and that it can
serve as the IR in a compiler's optimization stage.
The goal is not to outperform mature compilers like LLVM or GCC, as it
would require engineering effort far beyond the scope of this article.
We evaluate the RVSDG by generated code performance and size,
compilation time, and representational overhead.

\subsection{Implementation}\label{sec:implementation}

We have implemented \emph{jlm}, a publicly available~\cite{jlm} prototype compiler that uses the
RVSDG for optimizations. Its compilation pipeline is outlined in Figure~\ref{fig:jlmpipeline}. Jlm
takes LLVM IR as input, constructs an RVSDG, transforms and optimizes this RVSDG, and destructs it
again to LLVM IR. The SSA form of the input is destructed before RVSDG construction proceeds with
Inter- and Intra-PT. This additional step is required due to the control flow restructuring phase
of Intra-PT. Destruction discovers control flow by employing SCFR before it constructs SSA form to
output LLVM IR. Jlm supports LLVM IR function, integer, floating point, pointer, array,
structure, and vector types as well as their corresponding operations. Exceptions
and intrinsic functions are currently unsupported. The compiler uses two distinct
states to model side-effects: one for memory accesses and one for
non-terminating loops. We implemented the following optimizations in addition to CNE
and DNE:

\begin{figure*}
	\subfloat[Jlm compilation pipeline.\label{fig:jlmpipeline}] {
		\includegraphics[scale=0.84]{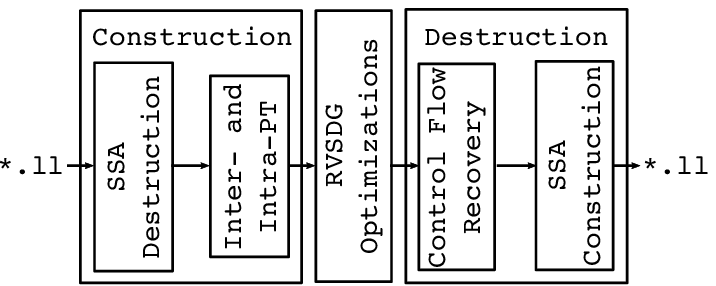}
	}
	\hfill
	\subfloat[Evaluation setup.\label{fig:evalsetup}]{
		\includegraphics[scale=0.84]{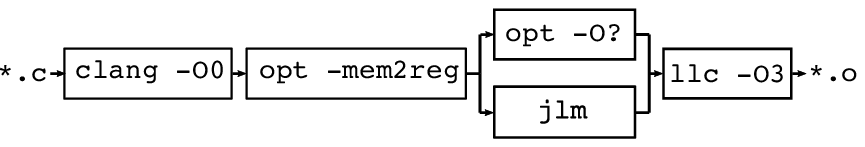}
	}
	\caption{Jlm's compilation pipeline and evaluation setup.}
	\label{fig:pipelines}
\end{figure*}

\begin{itemize}
	\item \textit{Inlining} (ILN): Simple function inlining.

	\item \textit{Invariant Value Redirection} (INV): Redirects invariant values from \thetam and
		\gammanodes.

	\item \textit{Node Push-out} (PSH): Moves all invariant nodes out of \gammam and \thetaregions.

	\item \textit{Node Pull-in} (PLL): Moves all nodes that are only used in one \gammaregion into
		the \gammanode. This ensures their conditional execution, while avoiding code bloat.

	\item \textit{Node Reduction} (RED): Performs simplifications, such as constant folding or
		strength reduction, similarly to LLVM's redundant instruction combinator
		(\texttt{-instcombine}), albeit by far not as many.

	\item \textit{Loop Unrolling} (URL): Unrolls all inner loops by a factor of four. Higher factors
		gave no significant performance improvements in return for the increased code size.

	\item \textit{$\theta-\gamma$ Inversion} (IVT): Inverts \gammam and \thetanodes where both nodes
		have the same predicate origin. This replaces the loop containing a conditional with a
		conditional that has a loop in its then-case.
\end{itemize}

We use the following optimization order: ILN INV RED DNE IVT INV DNE PSH INV DNE URL INV RED CNE
DNE PLL INV DNE.

\subsection{Evaluation Setup}

Figure~\ref{fig:evalsetup} outlines our evaluation setup. We use clang 7.0.1~\cite{clang} to
convert C files to LLVM IR, pre-optimize the IR with LLVM's \texttt{opt}, and then optimize
it either with \texttt{jlm}, or \texttt{opt} using different optimization levels. The optimized
output is converted to an object file with LLVM's \texttt{llc}. The pre-optimization step is
necessary to avoid a re-implementation of LLVM's \texttt{mem2reg} pass, since clang allocates all
values on the stack by default due to LLVM IR only supporting CFGs in SSA form.

We use the polybench 4.2.1 beta benchmark suite~\cite{polybench} to evaluate the RVSDG's
usability and efficacy. This benchmark suite provides structurally small benchmarks, and therefore
reduces the implementation effort for the construction and destruction phases, as well as the
number and complexity of optimizations.

The experiments are performed on an Intel Xeon E5-2695v4 running CentOS 7.4. The
core frequency is pinned to 2.0 GHz to avoid performance variations and thermal throttling effects.
All outputs of the benchmark runs are verified to equal the corresponding outputs of the
executables produced by clang.

\subsection{Performance}

\begin{figure}
	\centering
	\includegraphics[width=\linewidth]{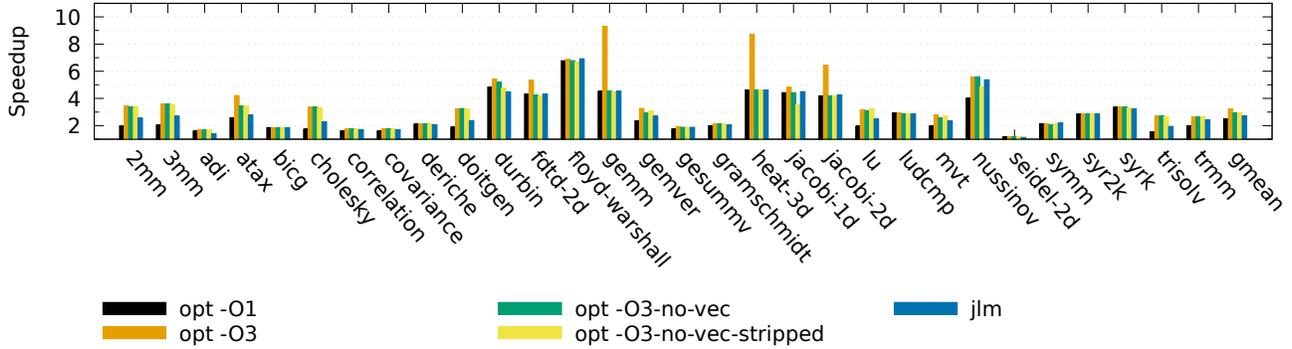}
	\caption{Speedup relative to \texttt{O0} at different optimization levels.}
	\label{fig:speedup}
\end{figure}

Figure~\ref{fig:speedup} shows the speedup at five different optimization levels. The \Ozero
optimization level serves as baseline. The \Othreenovec optimization level is the same as
\Othree, but without slp- and loop-vectorization. Optimization level
\Othreenovecstripped is the same as \Othreenovec, but the IR is stripped of named
metadata and attribute groups before invoking \texttt{llc}. Since jlm does not support metadata and
attributes yet, this optimization level permits us to compare the pure optimized IR against jlm
without the optimizer providing hints to \texttt{llc}. We omit optimization level \Otwo as
it was very similar to \Othree. The gmean column in Figure~\ref{fig:speedup} shows the
geometric mean of all benchmarks.

The results show that the executables produced by jlm (gmean 2.70) are faster than \Oone
(gmean 2.49), but slower than \Othree (gmean 3.22), \Othreenovec (gmean 2.95), and
\Othreenovecstripped (gmean 2.91). Optimization level \Othree attempts to vectorize
twenty
benchmarks, but only produces measurable results for eight of them: atax, durbin,
fdtd-2d, gemm, gemver, heat-3d, jacobi-1d, and jacobi-2d. Jlm would require a vectorizer to
achieve such speedups.

Disabling vectorization with \Othreenovec and \Othreenovecstripped shows that jlm
achieves similar speedups for fdtd-2d, gemm, heat-3d, javobi-1d, and jacobi-2d. The metadata
transferred between the optimizer and \texttt{llc} only makes a significant difference for durbin,
floyd-warshall, gesummv, jacobi-1d, and nussinov. In the case of gesummv and jacobi-1d, performance
drops below jlm.
Jlm is outperformed by optimization level \Oone at four benchmarks: adi, durbin, seidel-2d, and
syrk. We inspected the output files and found the following causes:

\begin{itemize}
	\item \textit{adi}: Jlm fails to eliminate load instructions from the two innermost loops. These
		loads have loop-carried dependencies with a distance of one to store instructions in the same
		loop, and can be eliminated by propagating the stored value to the users of the load's
		output. The corresponding LLVM pass is loop load elimination
		(\texttt{-loop-load-elim}). Jlm performance equals \Oone
        if this transformation is performed by hand.

	\item \textit{durbin}: Jlm fails to transform a loop that copies values between arrays to a
		\texttt{memcpy} intrinsic. This impedes LLVM's code generator to produce better code. The LLVM
		pass responsible for this transformation is the loop-idiom pass (\texttt{-loop-idiom}). If the
		loop is replaced with a call to \texttt{memcpy}, then jlm is better than \Oone.

	\item \textit{seidel-2d}: Similarly to adi, jlm fails to eliminate load instructions from
		the innermost loop. If the load elimination is performed by hand, then jlm achieves the
		same performance as \Oone.

	\item \textit{syrk}: Jlm fails to satisfactorily apply CNE due to an overly strict
		sequentialization of load and store instructions. Loads from the same address are not detected
		as congruent due to different state edge origins. An alias analysis pass would resolve this
		problem.
\end{itemize}

Figure~\ref{fig:speedup} shows that it is feasible to produce competitive code using the RVSDG,
but also that more optimizations and analyses are required in order to
reliably do so. Performance differences are not caused by inherent RVSDG characteristics,
but can be attributed to missing analyses, optimizations, and heuristics for their
application.
The above results and Table~\ref{tab:llvm-opts} indicate that an alias analysis
pass is particluarly required.

\subsection{Code Size}

\begin{figure}
	\centering
	\includegraphics[width=\linewidth]{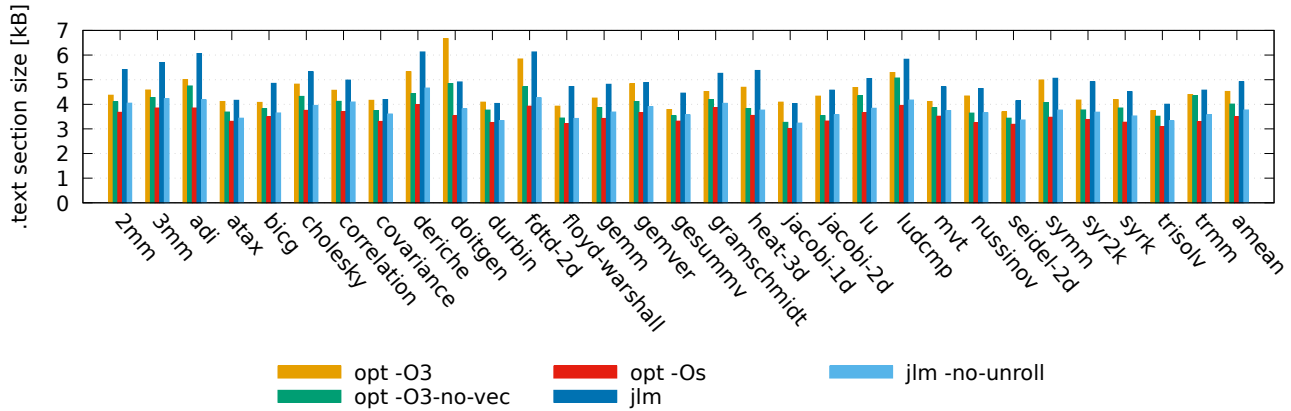}
	\caption{Code size at different optimization levels.}
	\label{fig:codesize}
\end{figure}

Figure~\ref{fig:codesize} shows the code size for \Othree, \Othreenovec, \Os,
and for jlm with and without loop unrolling. The amean column shows the arithmetic mean of all
benchmarks.
Optimization level \Othree produces on average text sections that are 11\% bigger than
\Othreenovec. Vectorization often requires loop transformations to make
loops amenable to the vectorizer, and the insertion of pre- and post-loop code. This affects code
size negatively, but can result in better performance.
The results also show that \Os produces smaller text sections than
\Othreenovec. This is due to more conservative optimization heuristics and the omission of
optimizations, \eg, aggressive instruction combination (\texttt{-aggressive-instcombine}) or
the promotion of by-reference arguments to scalars (\texttt{-argpromotion}).

Jlm produces ca. 39\% bigger text sections compared to \Os. The results
without loop unrolling show that this can be attributed to the naive heuristic used. Jlm does not
take code size into account and unrolls every inner loop unconditionally
four times, leading to excessive code expansion. Avoiding unrolling completely results in text
sections that are on average between \Othreenovec and \Os. This indicates that the
excessive code size is due to naive heuristics and shortcomings in the implementation, but not to
inherent characteristics of the RVSDG.

\subsection{Compilation Overhead}

\begin{figure*}
	\subfloat[Representational overhead.\label{fig:repoverhead}] {
		\includegraphics[scale=0.89]{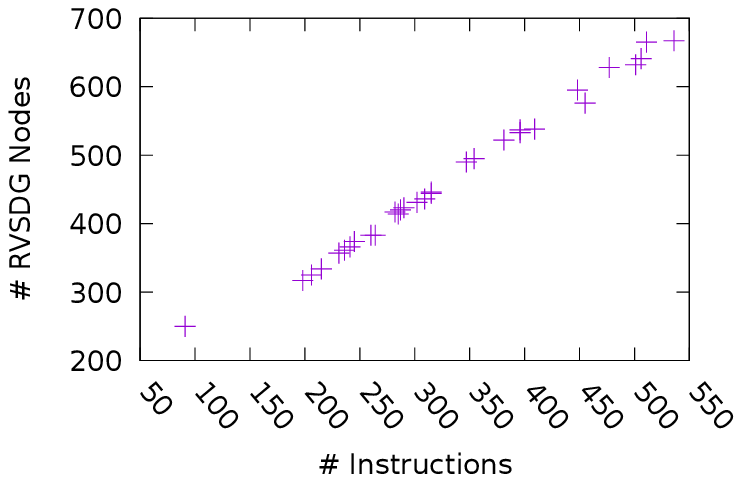}
	}
	\hfill
	\subfloat[Compilation times.\label{fig:timeoverhead}] {
		\includegraphics[scale=0.89]{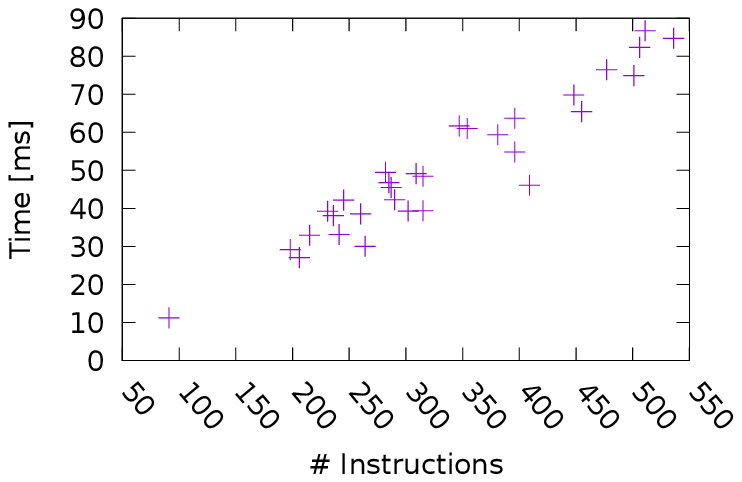}
	}
	\caption{Compilation overhead of jlm.}
	\label{fig:overhead}
\end{figure*}

Figure~\ref{fig:overhead} shows overhead in terms of IR size and time, with
Figure \ref{fig:repoverhead} relating LLVM instruction count to number of
RVSDG nodes, and Figure \ref{fig:timeoverhead} relating it to time spent on
RVSDG translation and optimizations.

Figure~\ref{fig:repoverhead} shows a clear linear relationship for all cases, confirming the
observations by Bahmann~\etal~\cite{Bahmann15} that the RVSDG is feasible in terms of space
requirements. Figure~\ref{fig:timeoverhead} also indicates a linear dependency, but with larger
variations for similar input sizes. We attribute this variation to the fact that construction,
optimizations, and destruction are also compounded by input structure. Structural
differences in the inter-procedure and control flow graphs lead to runtime variations in RVSDG
construction and destruction, as well as different runtimes for optimizations. For example, the
presence of loops in a translation unit determines whether loop unrolling is performed, while their
absence incurs no runtime overhead.
Figure~\ref{fig:overhead} shows that the RVSDG is feasible as an IR for optimizing
compilers in terms of compilation overhead.

\subsection{Comparison to LLVM}\label{sec:llvm-comparison}

LLVM 7.0.1 invokes 85 different analyses and optimization passes at optimization level \texttt{O3}.
Many of these passes are repeatedly invoked, resulting in a total of 266 invocations.
Section~\ref{sec:rep-invariants} already highlighted that 57, or 21\%, of these invocations are
from six helper passes that only establish invariants and detect structures necessary for
optimizations. Table~\ref{tab:llvm-opts} in Section~\ref{sec:rep-invariants} also shows that
LLVM necessitates SSA restoration in
fourteen optimization passes.
Because LLVM's CFG representation does not maintain necessary invariants and structures,
it requires (re-)computation of their information, which leads to a high number of
helper pass invocations.
This can
be observed in LLVM's optimization pipeline for loop optimizations:
... \emph{-loops -loop-simplify -lcssa-verification -lcssa} ... -loop-rotate -licm
-loop-unswitch \emph{-loop-simplify -lcsse-verification -lcssa} .... -loop-idiom -loop-deletion
-loop-unroll ... \emph{-loops -loop-simplify -lcssa-verification -lcssa} ... -loop-rotate
-loop-accesses ... . Depending on an optimization's position in the pipeline, several
helper passes must be executed before an optimization can be invoked. This is necessary to ensure
that the required information for an optimization is present, and up to date after 
invocation of other optimizations, \eg, jump threading (\texttt{-jump-threading}) or
CFG simplification (\texttt{-simplifycfg}). Thus, each added loop optimization
can necessitate several more helper passes in the optimization pipeline. A
similar pattern is seen with (basic) alias analysis (\texttt{-basicaa} and \texttt{-aa}),
which are invoked 37 times in total.

In contrast, the RVSDG establishes the necessary invariants and structures during construction,
and maintains them throughout compilation. The result is that jlm requires none of the
aforementioned helper passes and SSA restoration, \eg,
loop unrolling (URL) can be readily performed without the need to detect loops, their entry,
and exits. The cost is a more elaborate construction, which requires the
detection of the necessary information, and destruction, which requires the recovery
of control flow. However, RVSDG construction and destruction only need to be performed once
and, as demonstrated in Section~\ref{fig:overhead} and in Bahman~\etal~\cite{Bahmann15}, are
practically feasible.

\section{Conclusion}\label{sec:conclusion}

This paper presents a complete specification of the RVSDG IR for
an optimizing compiler. We provide construction and destruction algorithms, and show the RVSDG's
efficacy as an IR for analyses and optimizations by presenting Dead Node and Common Node
Elimination. We
implemented jlm, a publicly available~\cite{jlm} compiler that uses the RVSDG for optimizations, and
evaluate it in terms of performance, code size, compilation time, and representational overhead.
The results suggest that the RVSDG
combines the abstractions of data centric IRs with the CFG's advantages to optimize and generate
efficient control flow. This makes the RVSDG an appealing IR for optimizing compilers. A natural
direction for future work is to explore how features
such as exceptions can be efficiently mapped to the RVSDG. Another research direction would be to
extend the number of optimizations and their heuristics in jlm to a competitive level with CFG-based
compilers. This would provide further information about the number of necessary optimizations, their
complexity, and consequently the required engineering effort.

\bibliographystyle{plain}
\bibliography{defs,references}

\begin{thebibliography}{10}

\bibitem{Allen70}
Frances~E Allen.
\newblock Control flow analysis.
\newblock In {\em ACM Sigplan Notices}, volume~5, pages 1--19. ACM, 1970.

\bibitem{Alpern88}
B.~Alpern, M.~N. Wegman, and F.~K. Zadeck.
\newblock Detecting equality of variables in programs.
\newblock In {\em Proceedings of the {ACM} {SIGPLAN} Symposium on Principles of
  Programming Languages}, pages 1--11. ACM, 1988.

\bibitem{Bahmann15}
Helge Bahmann, Nico Reissmann, Magnus Jahre, and Jan~Christian Meyer.
\newblock Perfect reconstructability of control flow from demand dependence
  graphs.
\newblock {\em {ACM} Transactions on Architecture and Code Optimization},
  11(4):66:1--66:25, 2015.

\bibitem{Baxter89}
W.~Baxter and H.~R. Bauer, III.
\newblock The program dependence graph and vectorization.
\newblock In {\em Proceedings of the {ACM} {SIGPLAN} Symposium on Principles of
  Programming Languages}, pages 1--11. ACM, 1989.

\bibitem{Campbell93}
Philip~L Campbell, Ksheerabdhi Krishna, and Robert~A Ballance.
\newblock {Refining and Defining the Program Dependence Web}.
\newblock Technical report, {University of New Mexico}, 1993.

\bibitem{Carter03}
Larry Carter, Jeanne Ferrante, and Clark Thomborson.
\newblock {Folklore Confirmed: Reducible Flow Graphs are Exponentially Larger}.
\newblock In {\em Proceedings of the {ACM} {SIGPLAN} Symposium on Principles of
  Programming Languages}, volume~38, pages 106--114. ACM, 2003.

\bibitem{Castrillon13}
Jeronimo Castrillon.
\newblock {\em Programming Heterogeneous MPSoCs: Tool Flows to Close the
  Software Productivity Gap}.
\newblock PhD thesis, RWTH Aachen Univeristy, 2013.

\bibitem{Choi96}
Jong-Deok Choi, Vivek Sarkar, and Edith Schonberg.
\newblock Incremental computation of static single assignment form.
\newblock In {\em Proceedings of the International Conference on Compiler
  Construction}, pages 223--237. Springer-Verlag, 1996.

\bibitem{clang}
Clang.
\newblock Clang: A c language family frontend for llvm.
\newblock \url{https://clang.llvm.org}, 2017.
\newblock Accessed: 2019-10-30.

\bibitem{gcc}
GNU~Compiler Collection.
\newblock \url{https://gcc.gnu.org/}, 2018.
\newblock Accesssed: 2019-08-05.

\bibitem{Cordes10}
Daniel Cordes, Peter Marwedel, and Arindam Mallik.
\newblock Automatic parallelization of embedded software using hierarchical
  task graphs and integer linear programming.
\newblock In {\em Proceedings of the {ACM}/{IEEE} International Conference on
  Hardware/Software Codesign and System Synthesis}, pages 267--276, 2010.

\bibitem{Cytron91}
Ron Cytron, Jeanne Ferrante, Barry~K. Rosen, Mark~N. Wegman, and F.~K Zadeck.
\newblock Efficiently computing static single assignment form and the control
  dependence graph.
\newblock Technical report, 1991.

\bibitem{Dennis80}
Jack~Bonnell Dennis.
\newblock Data flow supercomputers.
\newblock {\em Computer}, 13(11):48--56, 1980.

\bibitem{Ding14}
Shuhan Ding, John Earnest, and Soner \"{O}nder.
\newblock {Single Assignment Compiler, Single Assignment Architecture: Future
  Gated Single Assignment Form}.
\newblock In {\em Proceedings of the International Symposium on Code Generation
  and Optimization}. ACM, 2014.

\bibitem{Ferrante87}
Jeanne Ferrante, Karl~J. Ottenstein, and Joe~D. Warren.
\newblock The program dependence graph and its use in optimization.
\newblock {\em {ACM} Transactions on Programming Languages and Systems},
  9(3):319--349, 1987.

\bibitem{Havlak93}
Paul Havlak.
\newblock {Construction of Thinned Gated Single-Assignment Form}.
\newblock In {\em Proceedings of the International Workshop on Languages and
  Compilers for Parallel Computing-Revised Papers}, pages 477--499. Springer,
  1993.

\bibitem{Horwitz288}
S.~Horwitz, J.~Prins, and T.~Reps.
\newblock On the adequacy of program dependence graphs for representing
  programs.
\newblock In {\em Proceedings of the {ACM} {SIGPLAN} Symposium on Principles of
  Programming Languages}, pages 146--157. ACM, 1988.

\bibitem{Horwitz88}
S.~Horwitz, T.~Reps, and D.~Binkley.
\newblock Interprocedural slicing using dependence graphs.
\newblock volume~23, pages 35--46. ACM, 1988.

\bibitem{Johnson03}
Neil Johnson and Alan Mycroft.
\newblock Combined code motion and register allocation using the value state
  dependence graph.
\newblock In {\em Proceedings of the International Conference on Compiler
  Construction}, pages 1--16. Springer-Verlag, 2003.

\bibitem{Johnson04}
Neil~E. Johnson.
\newblock Code size optimization for embedded processors.
\newblock Technical report, University of Cambridge, Computer Laboratory, 2004.

\bibitem{Johnson94}
Richard Johnson, David Pearson, and Keshav Pingali.
\newblock The program structure tree: Computing control regions in linear time.
\newblock pages 171--185. ACM, 1994.

\bibitem{Lattner04}
Chris Lattner and Vikram Adve.
\newblock {LLVM: A Compilation Framework for Lifelong Program Analysis \&
  Transformation}.
\newblock In {\em Proceedings of the International Symposium on Code Generation
  and Optimization}, 2004.

\bibitem{Lattner20}
Chris Lattner, Mehdi Amini, Uday Bondhugula, Albert Cohen, Andy Davis, Jacques
  Pienaar, River Riddle, Tatiana Shpeisman, Nicolas Vasilache, and Oleksandr
  Zinenko.
\newblock Mlir: A compiler infrastructure for the end of moore's law, 2020.

\bibitem{Lawrence07}
Alan~C. Lawrence.
\newblock {Optimizing compilation with the Value State Dependence Graph}.
\newblock Technical report, University of Cambridge, 2007.

\bibitem{llvm-bug1}
LLVM.
\newblock \url{https://bugs.llvm.org/show_bug.cgi?id=31851}, 2018.
\newblock Accesssed: 2018-05-07.

\bibitem{llvm-bug2}
LLVM.
\newblock \url{https://bugs.llvm.org/show_bug.cgi?id=37202}, 2018.
\newblock Accesssed: 2018-05-07.

\bibitem{llvm-bug3}
LLVM.
\newblock \url{https://bugs.llvm.org/show_bug.cgi?id=31183}, 2018.
\newblock Accesssed: 2018-05-07.

\bibitem{manjikian97}
Naraig Manjikian and Tarek~S Abdelrahman.
\newblock Fusion of loops for parallelism and locality.
\newblock {\em {IEEE} Transactions on Parallel and Distributed Systems},
  8(2):193--209, 1997.

\bibitem{Muchnick97}
Steven~S. Muchnick.
\newblock {\em Advanced Compiler Design and Implementation}.
\newblock Morgan Kaufmann, 1997.

\bibitem{Namballa04}
R.~Namballa, N.~Ranganathan, and A.~Ejnioui.
\newblock {Control and Data Flow Graph Extraction for High-Level Synthesis}.
\newblock In {\em IEEE Computer Society Annual Symposium on VLSI}, pages
  187--192, 2004.

\bibitem{Ottenstein90}
Karl~J. Ottenstein, Robert~A. Ballance, and Arthur~B. MacCabe.
\newblock {The Program Dependence Web: A Representation Supporting Control-,
  Data-, and Demand-driven Interpretation of Imperative Languages}.
\newblock In {\em Proceedings of the {ACM} {SIGPLAN} Conference on Programming
  Language Design and Implementation}, pages 257--271. ACM, 1990.

\bibitem{Ottoni05}
Guilherme Ottoni, Ram Rangan, Adam Stoler, and David~I. August.
\newblock {Automatic Thread Extraction with Decoupled Software Pipelining}.
\newblock In {\em Proceedings of the {ACM/IEEE} International Symposium on
  Microarchitecture}, pages 105--118. IEEE, 2005.

\bibitem{polybench}
Louis-No\"{e}l Pouchet.
\newblock Polybench/c 4.2.
\newblock \url{http://web.cse.ohio-state.edu/~pouchet.2/software/polybench/},
  2017.
\newblock Accessed: 2019-11-11.

\bibitem{Reissmann12}
Nico Reissmann.
\newblock Utilizing the value state dependence graph for haskell.
\newblock 2012.

\bibitem{jlm}
Nico Reissmann.
\newblock jlm.
\newblock \url{https://github.com/phate/jlm}, 2017.
\newblock Accessed: 2017-12-13.

\bibitem{Reissmann16}
Nico Reissmann, Thomas~L. Falch, Benjamin~A. Bj{\o}rnseth, Helge Bahmann,
  Jan~Christian Meyer, and Magnus Jahre.
\newblock Efficient control flow restructuring for {GPU}s.
\newblock In {\em Proceedings of the International Conference on High
  Performance Computing and Simulation}, pages 48--57, 2016.

\bibitem{Rosen88}
B.~K. Rosen, M.~N. Wegman, and F.~K. Zadeck.
\newblock Global value numbers and redundant computations.
\newblock In {\em Proceedings of the {ACM} {SIGPLAN} Symposium on Principles of
  Programming Languages}, pages 12--27. ACM, 1988.

\bibitem{Rugina00}
Radu Rugina and Martin~C. Rinard.
\newblock Recursion unrolling for divide and conquer programs.
\newblock In {\em Proceedings of the International Workshop on Languages and
  Compilers for Parallel Computing-Revised Papers}, pages 34--48.
  Springer-Verlag, 2001.

\bibitem{Sarkar91}
V.~Sarkar.
\newblock {Automatic Partitioning of a Program Dependence Graph into Parallel
  Tasks}.
\newblock {\em IBM Journal of Research and Development}, 35(5-6):779--804,
  1991.

\bibitem{Sharir80}
M.~Sharir.
\newblock Structural analysis: A new approach to flow analysis in optimizing
  compilers.
\newblock {\em Computer Languages}, 5(3-4):141--153, 1980.

\bibitem{Stanier12}
James Stanier.
\newblock {\em Removing and Restoring Control Flow with the Value State
  Dependence Graph}.
\newblock PhD thesis, University of Sussex, 2012.

\bibitem{Stanier211}
James Stanier and Alan Lawrence.
\newblock The value state dependence graph revisited.
\newblock In {\em Proceedings of the Workshop on Intermediate Representations},
  pages 53--60, 2011.

\bibitem{Stanier13}
James Stanier and Des Watson.
\newblock Intermediate representations in imperative compilers: A survey.
\newblock {\em ACM Computing Surveys (CSUR)}, 45(3):26:1--26:27, 2013.

\bibitem{Tarjan72}
R.~Tarjan.
\newblock Depth-first search and linear graph algorithms.
\newblock {\em {SIAM} Journal on Computing}, 1(2):146--160, 1972.

\bibitem{Tu95}
Peng Tu and David Padua.
\newblock {Efficient Building and Placing of Gating Functions}.
\newblock In {\em Proceedings of the {ACM} {SIGPLAN} Conference on Programming
  Language Design and Implementation}, pages 47--55. ACM, 1995.

\bibitem{Wegman91}
Mark~N. Wegman and F.~Kenneth Zadeck.
\newblock Constant propagation with conditional branches.
\newblock {\em {ACM} Transactions on Programming Languages and Systems},
  13(2):181--210, 1991.

\bibitem{Weise94}
Daniel Weise, Roger~F. Crew, Michael Ernst, and Bjarne Steensgaard.
\newblock Value dependence graphs: Representation without taxation.
\newblock In {\em Proceedings of the {ACM} {SIGPLAN} Symposium on Principles of
  Programming Languages}, pages 297--310. ACM, 1994.

\bibitem{Zaidi15-2}
Ali~Mustafa Zaidi.
\newblock {Accelerating control-flow intensive code in spatial hardware}.
\newblock Technical report, University of Cambridge, 2015.

\bibitem{Zaidi15}
Ali~Mustafa Zaidi and David Greaves.
\newblock Value state flow graph: A dataflow compiler ir for accelerating
  control-intensive code in spatial hardware.
\newblock {\em {ACM} Transactions on Reconfigurable Technology and Systems},
  9(2):14:1--14:22, 2015.

\bibitem{Zhang04}
F.~Zhang and E.H. D'Hollander.
\newblock Using hammock graphs to structure programs.
\newblock {\em IEEE Transactions on Software Engineering}, 30(4):231--245,
  2004.

\end{thebibliography}

\end{document}